\documentclass[twocolumn,aps]{revtex4}
\usepackage{graphicx}
\usepackage{amssymb}
\usepackage{amsmath}
\usepackage[usenames]{color}
\usepackage{latexsym}
\newcommand\be{\begin{equation}}
\newcommand\bea{\begin{eqnarray}}
\newcommand\eea{\end{eqnarray}}
\newcommand\ee{\end{equation}}
\newcommand\lD{\lambda_D}
\newcommand\lB{l_B}

\begin{document}


\title{Ion Induced Lamellar-Lamellar Phase Transition in
  Charged Surfactant Systems}
\author{Daniel Harries, Rudi Podgornik$^\dagger$ and V. Adrian Parsegian}
\affiliation{ Laboratory of Physical and Structural Biology,
National Institute of Child Health and Human Development, National
Institutes of Health, Bethesda, Maryland 20892-0924, USA \\
and \\
$^\dagger$ Faculty of Mathematics and Physics,  University of Ljubljana,
 and Department of Theoretical Physics, J. Stefan Institute, Ljubljana, Slovenia}

\author{Etay Mar-Or and David Andelman}
\affiliation{ School of Physics and Astronomy, Raymond and Beverly
Sackler Faculty of Exact Sciences, Tel Aviv University, Tel Aviv
69978, Israel}


\begin{abstract}

We propose a model for the liquid-liquid (${\rm L}_{\alpha}
\rightarrow {\rm L}_{\alpha'}$) phase transition observed in osmotic
pressure measurements of certain charged lamellae-forming
amphiphiles. The model free energy combines mean-field electrostatic
and phenomenological non-electrostatic interactions, while the
number of dissociated counterions is treated as a variable degree of
freedom that is determined self-consistently. The model,
therefore, joins two well-known theories: the Poisson-Boltzmann
theory for ionic solutions between charged lamellae, and
Langmuir-Frumkin-Davies adsorption isotherm modified to account for charged
adsorbing species. Minimizing the appropriate free energy for each
interlamellar spacing, we find the ionic density profiles and the
resulting osmotic pressure. While in the simple Poisson-Boltzmann
theory the osmotic pressure isotherms are always smooth, we observe
a discontinuous liquid\---liquid phase transition when
Poisson-Boltzmann theory is self-consistently augmented by
Langmuir-Frumkin-Davies adsorption. This phase transition depends on
the area per amphiphilic headgroup, as well as on non-electrostatic
interactions of the counterions with the lamellae, and interactions between
counterion-bound and counterion-dissociated surfactants. 
Coupling lateral phase transition in the
bilayer plane with electrostatic interactions in the bulk, our
results offer a qualitative explanation for the existence of the
${\rm L}_{\alpha} \rightarrow {\rm L}_{\alpha'}$ phase-transition of
DDABr (didodecyldimethylammonium bromide), but its apparent absence
for the chloride and the iodide homologues. More quantitative
comparisons with experiment require better understanding of the
microscopic basis of the phenomenological model parameters.

\end{abstract}

\maketitle

\newpage

\section{Introduction}

  From naturally occurring phospholipids to synthetic
double-chain surfactants, over a wide range of concentrations,
amphiphiles in aqueous solutions self-assemble into multilamellar
phases. The stability of the lamellar stack depends not only on the
type of amphiphile, but also on the competition between several
inter-lamellar interactions~\cite{parsegian93}. Attractive van der
Waals interactions are balanced by repulsive interactions. Hydration
repulsion usually dominates when the intervening water layer
spacings are small (typically $\lesssim 1$nm) or intermediate, while
electrostatic and `steric' undulation interactions usually prevail
at intermediate to large spacing, up to hundreds of
nanometers~\cite{helfrich78,evans86,andelman95,petrache98,zemb92a}.

Between charged surfactants, the stabilizing repulsion is typically
provided by the strong Coulomb interaction mediated by dissolved
counterions and salt~\cite{LesHouches}. These interactions are
particularly strong for lamellar-forming charged surfactants whose
counterions fully dissociate into solution. Salt can attenuate such
electrostatic interactions via ionic screening. For surfactants that
form flexible layers, the complicated yet important coupling between
layer elasticity, undulations, and electrostatic interactions must
also be considered~\cite{andelman95,safinya89,guttman93,harries03}.
But even when the effects of layer flexibility can be ignored,
electrostatic interactions in multilamellar charged systems are
non-trivial and, in general, difficult to understand because of the
intimate link between counterion dissociation, ionic screening, and
ion-specific non-electrostatic
interactions~\cite{parsegian71,arnold95,may02}.

The determination of the collapse pressure of membrane stacks by
Dubois et al.~\cite{zemb98} adds a new twist.  Synthetic cationic
double-chain surfactant, didodecyldimethylammonium (DDA$^+$) with
bromide as counterion (DDABr), is used to form a thermodynamically
stable lamellar phase.  The system undergoes a phase transition from
a swollen liquid-like (L$_\alpha$) lamellar phase to another, more
condensed, liquid-like lamellar phase (L$_{\alpha'}$). This phase
transition is induced by externally applied osmotic pressure; it is
seen as a plateau in the osmotic pressure versus inter-lamellar
spacing isotherms. Measured by small angle X-ray scattering, the
abrupt change in spacing is typically between $~$10\AA~to $~$100\AA.
In contrast, for the same surfactant with the bromide counterion
replaced by chloride, DDACl, there is no evidence of a first-order
transition~\cite{zemb98}. In fact, the experimental isotherm can be
well fit by the usual Poisson-Boltzmann (PB) theory
\cite{andelman95}. Further, with an iodide counterion the stack
made of DDAI surfactant remains collapsed and does not swell at
all~\cite{zembpc}. Remarkably, a discontinuous {\em increase} in area
per surfactant with no discernible in-plane positional order was
experimentally found to coincide with the collapse in bilayer
spacing. Clearly, this collapse is strongly coupled to a lateral
rearrangement in the bilayer plane, and cannot be solely the result
of neutralizing surfactant headgroups by their counterions.

It should perhaps come as no surprise that different halide
counterions interact differently with the charged DDA$^+$ surfactant
layers. The number of electrons, hence properties like
polarizability, vary widely for these ions, and we expect that
ion-membrane interactions will be different too. By ranking ions
according to their efficiency in salting-out proteins from solution,
Hofmeister was first to observe --- over a century ago --- that
different ions partition differently at aqueous interfaces
\cite{hofmeister}. The Hofmeister ranking is surprisingly
insensitive to the details of the interface
\cite{washabaugh85,ninham04,garrett04}. Often, however, the
preferential interaction follows the size and polarizability of the
ion; large ions tend to be less repelled from (or more attracted to)
oily interfaces.

It has been proposed that the added van der Waals attraction of the
ions to the higher index of refraction material may explain the
Hofmeister ranking
\cite{ninham97,attard88,ninham03,tobias02,gurau04}. Water ordering
around ions at the interface that is structured differently from the
bulk can also discriminate between ions. More polarizable ions, for
example, may be attracted to ordered water molecules at the
interface because of favorable interaction between dipoles and
induced-dipoles. Due to their amphiphilic, liquid-like nature,
surfactants present a special and complex interface to water and
salt ions. However, using measurements such as electrophoretic
mobility, nuclear magnetic resonance (NMR), and buoyancy
density-matching, a Hofmeister-like ranking of anions has also
emerged for ions at lipid interfaces
\cite{tatulian83,rydall92,petrache05}. Also, for single chain
micelle-forming cationic surfactants, the area per molecule follows
the Hofmeister series, increasing more in the presence 
of larger ions.~\cite{zemb04}.

As is evident from NMR experiments, different ions not only
associate differently with the amphiphile-water interface, but their
binding may also restructure the interface itself \cite{rydall92}.
Computer simulations indicate that the restructuring of the
amphiphilic headgroup region should be strongly influenced by the
counter-ion size \cite{sachs03}. Such conformational changes at the
interface are possible sources of non-ideal lipid mixing, because
ion binding at the interface may effectively create two incompatible
types of lipids: ion-bound and ion-detached. For example, both
experiments and simulations of {\it lipid mixtures}
\cite{feigenson93a,feigenson93b,blume00,noro99,nardi98,pabst05} show
that charged and uncharged lipids tend to demix so as to minimize
the line tension between the different mismatched lipid species,
often leading to lipid lateral phase separation in the membrane
plane. Perhaps most compelling are the phase transitions from
lamellar to inverted hexagonal phases of pure DOPS induced by
varying pH that changes the fraction of charged to uncharged
ionizable lipids \cite{rand03}.


Can an added non-electrostatic attraction of ions to the lipid-water
interface explain the observed transition for Br$^-$ ions? The
charge regulation model of Ninham and Parsegian \cite{parsegian71}
indicates that while an added attraction can significantly modify
pressure isotherms, it cannot account for a first-order phase
transition.

Here, we propose a phenomenological model that explains the
first-order phase transition in terms of an added coupling between
electrostatics and non-electrostatic specific interactions at the
interface. The model is motivated by the experimentally observed
lamellar-lamellar phase transition in charged surfactant
systems~\cite{zemb98} and is a relatively simple extension of
Poisson-Boltzmann theory.

The gist of our model is to consider the possibility that a fraction
of the counterions are not dissociated from the lamellar-forming
cationic DDA$^+$ surfactant, but rather stay associated with it on
the membrane plane to form a neutral complex. The degree of
dissociation is taken as a variational parameter in our free-energy
formulation, and is optimized for each inter-lamellar distance
\cite{parsegian71}. Further, we consider each lamella as composed of
a binary mixture of neutral (associated counterions) and charged
(dissociated counterions) surfactant species. Assuming an effective
attractive 2nd order virial coefficient between the two species, we
find possible lateral phase separation in the lamellar plane forming
neutral-surfactant rich and charged-surfactant rich phases, much
as in regular solution theory ~\cite{hill60,may02}. Like any
phenomenological model, our model relies on several parameters whose
exact molecular origin is not well known at present. However, using
reasonable values of these parameters, we are able to fit well the
experimental data.


Our model couples the Poisson-Boltzmann theory for the counterions
in solution with the Langmuir-Frumkin-Davies adsorption model that
regulates the amount of dissociated
counterions~\cite{hill60,adamson90,davis58,andelman96,andelman01}.
It is, therefore, an extension of the charge regulation model of
Ninham and Parsegian \cite{parsegian71}. Analogous coupling between
surface transitions and bulk interactions has been analyzed in the
context of hydration forces \cite{podgornik89,kornishev92} as well
as electrostatic interactions \cite{podgornik95}.

In the model, the differences between monovalent ions 
(Cl$^-$, Br$^-$, and I$^-$) are accounted for by using different 
interaction parameters
between ions and lipid interface and between ion-bound and
ion-detached lipids. Together with the repulsive hydration force,
known to act strongly at small separations such as those found in 
the collapsed
phase, these interaction parameters are sufficient to reproduce the
experimental observations. Our main emergent result supports a
lamellar-lamellar phase-transition as function of externally applied
osmotic pressure for ions such as Br$^-$. At small osmotic pressure
and large inter-lamellar distances, most of the Br$^-$ ions are
dissociated, and the isotherm follows the PB
result~\cite{andelman96,andelman01}. However, for larger pressures
and smaller separations, a large fraction of the Br$^-$ ions remain
associated, causing a lateral phase transition. Because of the
coupling between electrostatics and the entropy of ions in solution, the
lateral phase transition also leads to the discontinuous jump in
inter-lamellar spacing witnessed in the osmotic pressure isotherm.


We further consider the effect of added salt on the equilibrium state of the
system. The main effect of salt is to screen electrostatic
interactions, to reduce the coupling between the layers, and thus to
diminish the magnitude of the first-order transition
jump~\cite{warr96}. Indeed, our model predicts that
for more than a critical amount of salt the phase transition
disappears altogether.

The outline of the paper is as follows: in Sec. II we present an
extension to the usual PB theory, taking into account the
non-electrostatic degrees of freedom and treating separately the
counterion-only and added-salt cases. In Sec. III we present our
numerically calculated isotherms to show the possibility of a
lamellar-lamellar phase transition. We then discuss the
link with experiments and comment on ion-specific effects.
In Sec. IV we discuss our findings and end in Sec. V
with a short summary and remarks on
possible future directions.


\section{Extended Poisson-Boltzmann Theory}

\subsection{Model}

The lamellar stack is composed of  bilayers of double-chain
surfactants such as DDA separated by regions of aqueous solution.
The charged surfactant hydrophilic headgroups point towards the
water region, while the hydrophobic tails are packed in the inner
lamellar region, away from the polar water environment; for DDABr,
the thickness of the hydrocarbon part of the bilayer is of the order
of 24--26\AA. The lamellar stack can be modeled as a one-dimensional
periodic system. This approximates the lamellar lateral extent as
infinite and each lamella as perfectly planar and rigid. We consider
only the unit cell of the lamellar stack, as is depicted in
Figure~1.
\begin{figure}
  \includegraphics[width=70mm,clip,trim= 0 0 0 0]{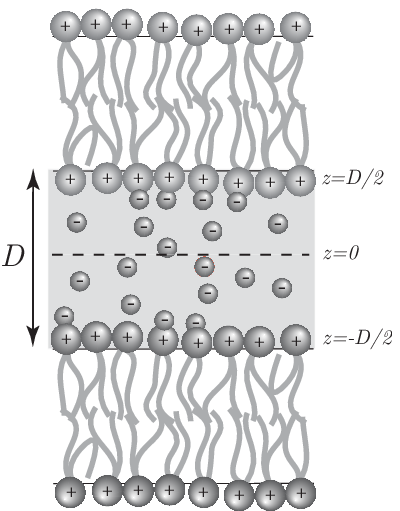}
  \caption{\footnotesize\textsf{A schematic view of the lamellar
      phase. The unit
  cell width includes only the water region,
  $-D/2<z<D/2$, and is
  composed of two charged surfaces located at $z=\pm D/2$
  (each being one leaflet of the surfactant bilayer). The
  counterions can adsorb on the two surfaces  or dissociate into the
  inter-membrane water region of thickness $D$.
  }}
\end{figure}

 For convenience, the unit cell width $D$ spans only the aqueous
inter-lamellar region, while the periodicity of the lamellar stack
includes also the bilayer thickness: $D+D_m$. All local quantities
depend only on the perpendicular coordinate, $z$. Because the
solubility of single surfactants (like DDA) in water is extremely
low, we assume that all surfactants reside within the lamellar
bilayers. These bilayers are the two bounding interfaces of the unit
cell.

The model system, hence, is composed of two charged interfaces
separated by a distance $D$. The system is overall electroneutral;
the amount of counterions in the aqueous region is exactly balanced
by the amount of charged surfactant on the two interfaces.
Note that because of the insolubility assumption, the ratio between
the lamellar bilayer thickness and $D$ uniquely determines the
relative concentration of surfactant and water.

The midplane of the unit cell, see Figure 1, is chosen at $z=0$,
from which the two interfaces bearing equal charge densities are
located at $z=\pm D/2$. Because the system is symmetric about the
$z=0$ midplane, it is enough to consider a unit cell in the range
$0\leq z\leq D/2$. The counterion concentration (mole per liter)
$c(z)$ and the mean-field electrostatic potential $\psi(z)$ depend
on the perpendicular coordinate $z$. All variables calculated at the
lamellar surface, $z=D/2$, will be denoted by a subscript $s$ (e.g.,
$\psi_s$), while those calculated at the symmetric midplane, $z=0$,
by a subscript $m$ (e.g., $\psi_m$).

Our aim is to calculate the equation of state for this lamellar
symmetry and to express it as a relation between the thermodynamic
variables: inter-lamellar distance, $D$, and  osmotic pressure,
$\Pi$. Just as in the van der Waals phenomenological theory of phase
transitions, we search for thermodynamically stable states or stable
branches of the free energy. When we find more than one branch, we
can use the Maxwell construction to obtain the coexistence region.

Minimization of the model free energy with respect to electrostatic
and non-electrostatic degrees of freedom (see below) will eventually
lead to the equation of state, our final goal. The overall free
energy $F_v+F_s+F_{\rm hyd}$ per unit cell area is a sum of volume
contributions coming from the electrolyte solution within the cell,
including  ion (electrostatic) $F_v$ terms, surface contributions
having their origin at the interfaces $F_s$, and hydration
interaction $F_{\rm hyd}$, dominant at small $D$ separations.

Because we approximate the hydration interaction as a separable
term, we can independently minimize the electrostatic contribution
to the free energy. As we show next, these interactions already
suffice to account for a discontinues phase transition, while
$F_{\rm hyd}$ is needed only to account for the experimentally found
pressure at small $D$. We therefore discuss only the minimization of
$F_{\rm tot}=F_{v}+F_{s}$ below, and return to present and discuss
the added $F_{\rm hyd}$ when we compare theory and experiment in the
Results section.

Because the system is extensive in the interfacial area, the ion
(electrostatic) volume free energy per unit area, $F_v$,  is taken
as the appropriate intensive quantity
\be F_v=\int_0^{D/2}
\left[-ec\psi- \frac{\varepsilon}{8\pi}\left(\psi'\right)^2+
{k_BT}c\left(\ln\left(c/c_0\right)-1\right)\right]\,
\mathrm{d}z\label{e1}~.
\ee
The first two terms include the standard electrostatic energy, with
$\varepsilon\simeq 80$ the dielectric constant of water. The last
term is the entropy of mixing in the dilute limit, where $c_0$ is a
reference concentration, $T$ the temperature, and $k_B$ the Boltzmann
constant. Because all counterions in solution originate from a
surfactant molecule, their integrated concentration (per unit area)
must be equal in magnitude and opposite in sign to the surface
charge density
\be \sigma=-e\int^{D/2}_{0} c(z) \,\mathrm{d}z\label{e2}~. \ee
This is also a charge density condition and can be translated via
Gauss' law into the electrostatic boundary condition (in gaussian units):
$\psi'(D/2)=\psi'_s=4\pi\sigma/\varepsilon$, linking the surface
electric field $\psi'_s$ with the surface charge density $\sigma$.
Unlike the usual PB theory where either the surface charge or the
surface potential is held fixed, here $\sigma$ is a self-adjusting
parameter which will be determined variationally from minimizing the
total free energy $F_{\rm tot}$.

The second part of the total free energy comes from the surface free
energy contributions of the amphiphiles residing on the planar
bilayers. The surface free energy $F_{s}$ has electrostatic and
non-electrostatic energy terms as well as a lateral mixing entropy
contribution. Expressed in terms of the surface area
fraction $\eta_s=a^2\sigma/e$ of charged surfactants,
\bea a^2F_s&=&e\psi_s\eta_s
-\hat{\alpha}\eta_s-\frac{1}{2}\hat{\chi}\eta_s^2 \nonumber\\ 
&+&k_B T\left[\eta_s\ln\eta_s+(1-\eta_s)\ln(1-\eta_s)\right],
\label{e3}\eea
where the first term couples between the surface charge and surface
potential. The other terms are the enthalpy and entropy of a
two-component liquid mixture: charged surfactant with area fraction
$\eta_s$ and neutralized, ion-bound surfactants with area fraction
$1-\eta_s$. The parameters $\hat{\alpha}$ and $\hat{\chi}$ are
phenomenological, respectively describing the
counterion\---surfactant and the surfactant\---surfactant
interactions at the surface. As in the charge regulation model
\cite{parsegian71}, here $\hat{\alpha}<0$ means that there is an
added non-electrostatic attraction (favorable adsorption free
energy) between counterions and the surface; the more counterions
are associated at the surface, the smaller the amount of remaining
charged surfactant.

The parameter $\hat{\chi}$  is the most crucial and unique element
in our model. Representing non-ideal mixing tendencies in the
bilayer plane as in regular solution theory, it alone (together with
the usual components of standard PB theory) is sufficient to account
for a coupled transition in the bilayer plane and in the bulk. As in
the Frumkin adsorption model
\cite{hill60,adamson90,davis58,andelman96,andelman01}, a positive
$\hat{\chi}$ parameter represents the tendency of surfactants on the
surface to phase separate into domains of neutral and charged
surfactants.

Changing to dimensionless variables, we define $y(z)\equiv
e\psi(z)/k_BT$, $\phi(z)\equiv a^3c(z)$,
$\alpha\equiv\hat{\alpha}/k_B T$, $\chi=\hat{\chi}/k_B T$, and take
for convenience $c_0 a^3=1$. Then, the total free energy is written
as a functional of the variables $y(z)$, $\phi(z)$, and a function of $\eta_s$,
and includes the conservation condition, eq~\ref{e2}, via a
Lagrange multiplier, $\mu$:

\bea
 \frac{a^2}{k_B T}F_{\rm tot}[y,\phi;\eta_s]=\frac{a^2}{k_B
 T}F_v+\frac{a^2}{k_B T}F_s &&\nonumber \\-
 \mu\left[\eta_s-\frac{1}{a}\int_{0}^{D/2}\phi(z)\,{\rm d}z\right] &&\nonumber\\
 =\frac{1}{a}\int_0^{D/2}\left[-y(z)\phi(z)-\frac{a^3}{8\pi\lB}(y')^2+\phi(\ln\phi-1)\right]{\rm
 d}z &&\nonumber\\
 +~y_s\eta_s-\alpha\eta_s-\frac{1}{2}\chi\eta_s^2+\eta_s\ln\eta_s+(1-\eta_s)\ln(1-\eta_s)
 &&\nonumber\\
  -~\mu\left[\eta_s-\frac{1}{a}\int_0^{D/2}\phi(z)\,{\rm d}z\right] &
  & \label{e4}~.
\eea
Next, we minimize
$F_{\rm tot}$ with respect to the surface variable $\eta_s$, and the two
continuous fields $\phi(z)$, $y(z)$:\, ${\rm d}F_{\rm tot}/{\rm d} \eta_s=\delta
F_{\rm tot}/\delta \phi(z)  = \delta F_{\rm tot}/\delta y(z) =0,$
corresponding to three coupled equations of state
\bea
\frac{\eta_s}{1-\eta_s}&=&\exp\left(\mu+\alpha+\chi\eta_s-y_s\right)\label{e5}\\
\phi(z)&=&\exp\left(-\mu+y(z)\right)\label{e6}\\
y''(z)&=&\frac{4\pi e^2}{\varepsilon k_B T
a^3}\phi(z)=\frac{4\pi\lB}{a^3}\phi(z)\label{e7}.
\eea
 The free energy $F_{\rm tot}$ is also a function of surface
potential $y_s$ and of the inter-lamellar spacing $D$. The
differentiation with respect to $D$ gives the osmotic pressure (to
be discussed in Sec. II.B), while the variation with respect to
$y_s$ gives the usual electrostatic boundary condition:
\be
  y'(D/2)=y'_s=\frac{4\pi\eta_s}{a^2}\label{e7a}~,
\ee
where $\lB=e^2/(\varepsilon k_B T)$ is the Bjerrum length, equal to
about 7\AA\ at room temperature for aqueous solutions
($\varepsilon=80$). The Lagrange multiplier, $\mu$,
acts as a chemical potential, but with the
important difference that it is related not to the bulk reservoir
concentration, but rather to the concentration at the midplane,
$\phi_m$. For a single counterion type, we can choose, without loss
of generality, the potential at the mid-plane to be zero,
$y_m=e\psi_m/k_B T=0$ and then from eqs~\ref{e6}-\ref{e7},
\bea
\phi(z)&=&\phi_m{\rm e}^{y(z)}\nonumber\\
\phi_m&=&{\rm e}^{-\mu} \nonumber\\
y''(z)&=&\frac{4\pi \lB}{a^3}\phi_m{\rm e}^{y(z)} \label{e8}~.
\eea
Not surprisingly, we recover the Poisson-Boltzmann equation
\ref{e8} connecting the electrostatic potential $y(z)$ with the
counterion concentration $\phi(z)$ in the solution. This can be
 expected since the non-electrostatic contributions enter
only via the surface interactions expressed in eq~\ref{e5}.

Rewriting eq~\ref{e5}, we arrive at an expression similar to the
Langmuir-Frumkin-Davies adsorption isotherm~\cite{davis58},
\be
{\eta_s}=\frac{1}{1+\phi_m{\rm
e}^{-\alpha-\chi\eta_s+y_s}}\label{e9}~,
\ee
with the following modifications: the
concentration $\phi_m$ at the midplane replaces the bulk
concentration (because of the constraint of overall charge
neutrality), and the surface interaction energy in the exponent
contains the electrostatic part $y_s$  \cite{parsegian71}. The
simpler Langmuir isotherm is recovered in the limit of non-charged
surfaces, $y_s=0$, and no surface interaction, $\chi=0$:
\be
{\eta_s}=\frac{1}{1+\phi_m{\rm e}^{-\alpha}}\label{e10}~.
\ee
Here, there is a unique relation between $\phi_m$ and $\eta_s$,
while in the general case of non-zero $y_s$ and $\chi$, the
generalized Langmuir-Frumkin-Davies equation offers a transcendental
relation between $\eta_s$, $\phi_m$ and $y_s$ with the possibility
of more than one solution.

The solution of the PB equation \ref{e8} for two symmetric charged
surfaces separated by a distance $D$, each having a surface charge
density of $\sigma$, has a well-known analytic form
\cite{vap66,parsegian71,andelman95}
\be
y(z)=-\ln\left[\cos^2(Kz)\right]
\label{e11}~,
\ee
where the constant $K$ is determined from the electrostatic
boundary condition $y'(D/2)=y_s'=4\pi\lB\eta_s/a^2$,
eq~\ref{e7a}, as
\be
KD\tan(KD/2)=\frac{2\pi \lB}{ a^2}\eta_s D
\label{e12}~.
\ee
We can now express $y_s$ and $\phi_m$ as function of a
single dimensionless variable $u\equiv KD/2$:
\bea
y_s&=&-\ln\left[\cos^2(u)\right]\nonumber\\
\eta_s&=&\frac{C_1}{D}u\tan(u)\nonumber\\
\phi_m&=&\frac{C_2}{D^2}u^2 \label{e13}~,
\eea
with the constants $C_1$ and $C_2$ obtained as
\bea
C_1&=&\frac{a^2}{\pi \lB} \nonumber\\
C_2&=&\frac{2 a^3}{\pi \lB}=2aC_1 \label{e14}~.
\eea
The solution of the above equations, together with the adsorption
isotherm eq~\ref{e9}, completely determines the counterion
density profile and the mean electrostatic potential via the
solution $K = K(D, \alpha, \chi)$.

\subsection{Equation of State and the ${\rm L}_{\alpha}
\rightarrow {\rm L}_{\alpha'}$ Phase Transition}

We now solve the basic set of equations derived in the previous section.
Substituting $\phi_m$, $\eta_s$, and $y_s$ into the
Langmuir-Frumkin-Davies isotherm, eq~\ref{e9}, we get an equation for $u$
\be \frac{D}{C_1 u\tan(u)}=1+\frac{C_2}{\cos^2(u)}
\left(\frac{u}{D}\right)^2
\exp\left(-\alpha-\chi\frac{C_1}{D}u\tan(u)\right)\label{e15}~.
\ee
Obviously, for each imposed distance $D$, we can extract $u=KD/2$
from eq~\ref{e15}. Once $u$ is known, it can be
substituted into eq~\ref{e13}, wherefrom $\eta_s$, $\phi_m$, $y_s$
follow. Their values completely determine the
potential profile $\psi(z)=k_B T y(z)/e$ and counterion profile
$c(z)=\phi(z)/a^3$.

If there is only one solution for $u$, the system has one stable
state. Multiple $u$ solutions indicate the possibility of
coexistence between several phases, as well as transitions between
them. Each phase corresponds to a separate branch of the free energy
with its own dependence on $D$. If all solutions of eq~\ref{e15}
are non-zero, then we can have a first-order transition between two
stable phases.

The free energy as a function of $D$ and $\eta_s(D)$ can be obtained
by substituting the results of minimization back into
eq~\ref{e4}, yielding
\be
\frac{a^2}{k_B T}F_{\rm tot}=
-\frac{1}{2\pi\lB}\left[4\pi\lB{\eta_s}
-\frac{1}{2}(Ka)^2{D}\right]+\frac{1}{2}\chi
\eta_s^2+\ln(1-\eta_s) \label{e16}~.
\ee
The appropriate isotherm is now obtained by taking the derivative of
the free energy with respect to $D$, giving the osmotic pressure
$\Pi(D)$ as
\be \Pi(D)=-\frac{\rm d F_{\rm tot}}{{\rm d} D}\label{e17}~.\ee
As is usual in PB theory \cite{andelman95}, the osmotic pressure can
also be calculated from the {\it contact theorem}, which
relates the osmotic pressure with the value of the counterion
concentration at the interface. Counterion concentration at the
interface ($z=D/2$) can be, in turn, connected with the
concentration at the midplane ($z=0$), thus yielding  an
alternative form of the osmotic pressure as
\be \Pi(D) = k_{B}T~c_{m}. \label{e17a}\ee
This latter equation has exactly the same form as in the standard PB
theory. The only way non-electrostatic terms of the free energy
influence the osmotic pressure is via the solution of
eq~\ref{e15}. Both forms of the osmotic pressure,
eqs~\ref{e17} and \ref{e17a}, of course yield exactly the same
values.

A typical isotherm $\Pi(D)$ (in Pascal units) calculated using
eqs~\ref{e15} and \ref{e17a} is shown in Figure 2a, and the
corresponding surface charge density in Figure 2b. Note that is this
example we do not include contributions from hydration. We discuss
the parameter range in the next section, and choose the parameters here
 to be $\alpha=-6$, $a=8$\AA\ and $\chi=12$. The isotherm
clearly exhibits a first-order phase transition from one free energy
branch at large inter-lamellar separation $D$ to another at smaller
$D$, and with a coexistence region in between. For large values of
$D$, $D\ge 64$\AA, most counterions are dissociated from
surfactants, $\eta_s\lesssim 1$, and the osmotic pressure follows
the standard PB theory for (almost) fully dissociated surfactants.
For $D$ smaller than $39$\AA, $\eta_s\le 0.1$, and the isotherm
follows another branch, characterized by a much smaller surface
charge of only about $10 \%$ of the fully dissociated value. For
$39{\rm \AA}\le D\le 64$\AA\ the system exhibits a coexistence
between two phases.  The pressure has a plateau and $\eta_s$ changes
from one branch to the second.

\begin{figure}
\includegraphics[keepaspectratio=true,width=80mm,clip=true]{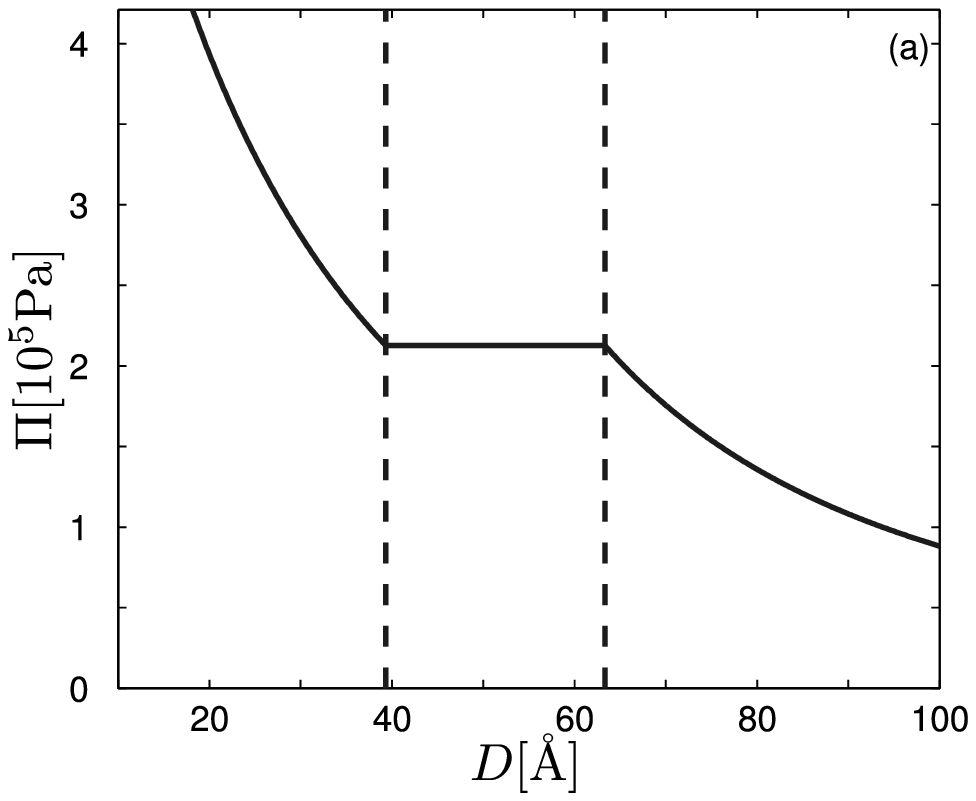}
\includegraphics[keepaspectratio=true,width=80mm,clip=true]{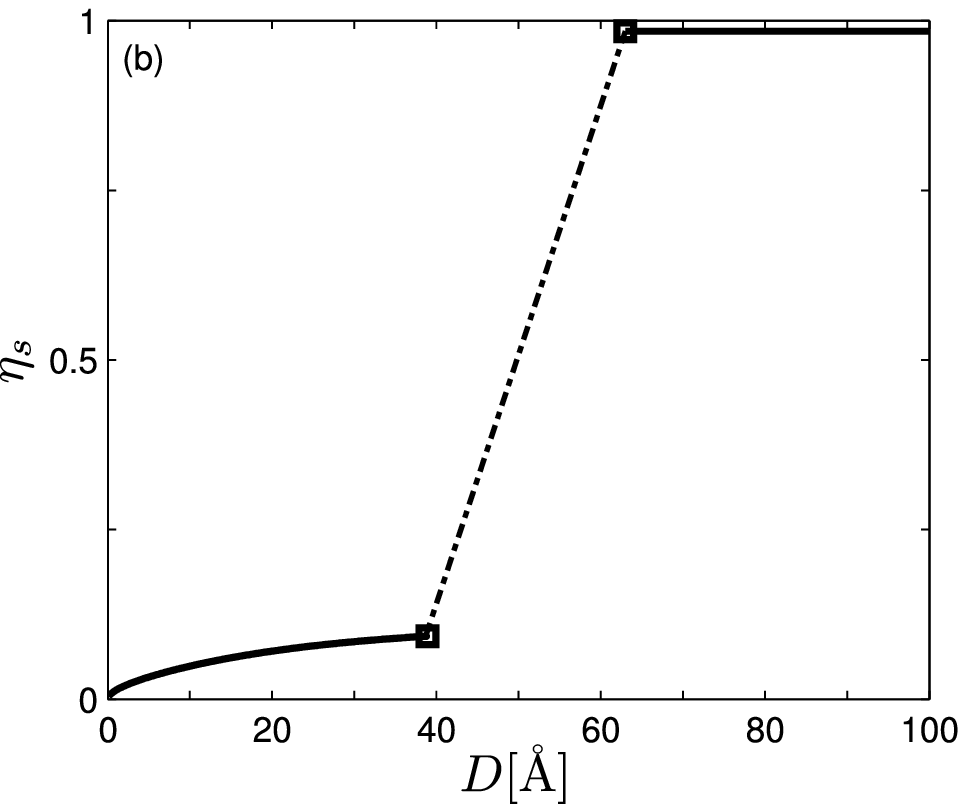}
  \caption{\footnotesize\textsf{
  (a) The osmotic pressure $\Pi$ in units of $10^5$\,Pascals
  ($\approx$1atm); and,
  (b) the area fraction $\eta_s=a^2\sigma /e$ of surface charges,
  as function of inter-lamellar spacing $D$ for $\alpha=-6$, $\chi=12$
  and $a=8$\,\AA.
  The Maxwell construction
  gives  a coexistence between a phase with $D\simeq 39$\,\AA\ and low 
$\eta_s\le 0.1$,
  and another with $D\simeq 64$\,\AA\ and  $\eta_s \approx 1$. In (b)
  the two coexisting phases are denoted by squares and the
  dotted-dashed 
line shows the tie-line
  in the coexistence region.}}
\end{figure}

 The plateau in the
osmotic pressure between the two stable solutions on Figure~2a is
evaluated using the usual Maxwell construction. This plateau in
$\Pi(D)$ indicates a first-order phase transition between the
solutions corresponding to two stable branches of the free energy: a
condensed one with $D\simeq 39$\,\AA\ and a dilute one with $D\simeq
64$\,\AA. Although it is hard to fit the phenomenological parameters
$\alpha$, $a$ and $\chi$ to the DDABr results of Ref. \cite{zemb98},
we believe that the mechanism proposed here and the typical results
presented in Figure 2 are relevant to the experimental system. More
details are given in Sec.~III below, where we show how the phase
transition depends on model parameters.

\subsection{Added-Salt Case}

After demonstrating the possibility of a phase transition for the
counterions only, we generalize the results by coupling the lamellar
system to a reservoir of 1:1 monovalent salt. This generalization is
easily achieved within our model. The main effect of the salt is to
diminish or even to eliminate completely the phase transition.

The main modifications of the model wrought by the introduction of salt ions
are as follows. First we must consider ionic concentration profiles
independently for positive and negative ions: $\phi^+(z)$ and $\phi^-$(z).
%
%
Note that we do not distinguish between the negative salt
counterions and those dissociated from the surfactant. The free
energy is written, similarly to eq~\ref{e4}, in terms of rescaled
variables with the introduction of two chemical potentials. The
first, $\mu^+$, is coupled to the excess amount of co-ions with
respect to the reservoir, stemming from the integral of $\phi^+$ in
the region between the two plates. The second, $\mu^-$, is coupled
with the excess number of counterions, stemming from the integral of
$\phi^-$ in the region between the two plates with the added
contribution of surface charges $1-\eta_s$. Thus, the total free
energy is given by
 \bea
 &~&\frac{a^2}{k_B T}F_{\rm tot}[y,\phi;\eta_s]=
 \frac{a^2}{k_B T}F_v+\frac{a^2}{k_B
 T}F_s \nonumber\\
 &+& \mu^+\left[\frac{1}{a}\int_{0}^{D/2}[\phi^+(z)-\phi_0^{+}]\,{\rm d}z\right] \nonumber\\
 &+&\mu^-\left[\frac{1}{a}\int_{0}^{D/2}[\phi^-(z)-\phi_0^{-}]\,{\rm d}z +(1-\eta_s)\right]
 \nonumber\\
&=&\frac{1}{a}\int_0^{D/2}\left[-y(z)\left[\phi^-(z)-\phi^+(z)\right]\right.\nonumber\\
&-&\left. \frac{a^3}{8\pi\lB}(y')^2+
\phi^-(\ln\phi^{-}-1)+\phi^+(\ln\phi^{+}-1)\right]{\rm d}z \nonumber\\
&+& y_s\eta_s-\alpha\eta_s-\frac{1}{2}\chi\eta_s^2+\eta_s\ln\eta_s+(1-\eta_s)\ln(1-\eta_s)\nonumber\\
&+&\mu^+\left[\frac{1}{a}\int_{0}^{D/2}[\phi^+(z)-\phi_0^{+}]\,{\rm
d}z\right] \nonumber\\
&+&\mu^-\left[\frac{1}{a}\int_{0}^{D/2}[\phi^-(z)-\phi_0^{-}]\,{\rm
d}z ~+~(1-\eta_s)\right]\label{e19}~. \eea
Taking now the variation of the above free energy with respect to $\eta_s$,
$\phi^\pm$ and $y$, we get the Euler-Lagrange equations,
\bea
\frac{\eta_s}{1-\eta_s}&=&\exp\left(\mu^{-}+\alpha+\chi\eta_s-y_s\right)\label{e20}\\
\phi^-(z)&=&\exp\left(-\mu^{-}+y(z)\right)\label{e21}\\
\phi^+(z)&=&\exp\left(-\mu^{+}-y(z)\right)\label{e22} \\
 y''(z)&=&\frac{4\pi e^2}{\varepsilon k_B T
 a^3}\left(\phi^-(z)-\phi^+(z)\right)\nonumber\\ &=&
 \frac{4\pi\lB}{a^3}\left(\phi^-(z)-\phi^{+}(z)\right)\label{e23}~.
 \eea
The final variation with respect to $y_s$ gives the usual
electrostatic boundary condition relating the surface electric field
with the surface charge density: $y_s'=(4\pi\lB/a^2)\eta_s$,
eq~\ref{e7a}. Requiring that the bulk concentration of co-ions
and counterions matches the reservoir concentration, $\phi_0$, where
the potential vanishes, $y=0$, it is easily verified from
eqs~\ref{e21}-\ref{e23} that
\bea
 \mu^\pm&=&-\ln \phi_0\label{e24}\\
\phi^\pm(z)&=&\phi_0{\rm e}^{\mp y(z)}\label{e25}\\
y''(z)&=&\frac{8\pi\lB\phi_0}{a^3}\sinh\left(y(z)\right)=\lD^{-2}\sinh
y \label{e26}~,
 \eea
where
 \be
 \lD=\left(\frac{8\pi\lB\phi_0}{a^3}\right)^{-1/2}\label{e27}~,
 \ee
is the Debye-H\"uckel screening length. Inserting eq~\ref{e24}
into \ref{e20} leads to a  Langmuir-Frumkin-Davis
isotherm, now of the form:
 \be
 {\eta_s}=\frac{1}{1+\phi_0{\rm
 e}^{-\alpha-\chi\eta_s+y_s}}\label{LFD1}~.
 \ee
This equation resembles eq~\ref{e9}, only that in
eq~\ref{LFD1} the reservoir concentration $\phi_0$ takes the
place of $\phi_m$. Moreover, the value of the potential at the
midplane is not fixed, but rather is determined from $\phi_0$, as
are the concentrations, $\phi_m^\pm=\phi_0\exp(\mp y_m)$.

The PB equation in presence of salt, eq~\ref{e26}, depends on the
electrostatic boundary conditions and has a well-known solution
expressed via an elliptic integral \cite{andelman95}. The first
integration of the PB equation, eq~\ref{e26}, from the midplane
position ($z=0$) to an arbitrary $z$ gives 
 \be
  y'(z)= \frac{1}{\lD}\sqrt{2\cosh y(z)-2\cosh y_{m}}
 \quad. \label{first_int}
 \ee
The boundary condition can be inserted in eq~\ref{first_int},
yielding
 \be
 \cosh y_{s}=\cosh y_{m} +\frac{\pi\lB}{a\phi_0}\eta_s^2
 \quad. \label{bc_2p12}
 \ee
A further integration can be written in terms of an elliptic
function 
 \be
 \frac{z}{\lD}=\int_{y_{m}}^{y}\frac{{\rm d}w}{\sqrt{2\cosh{w}
 -2\cosh{y_{m}}}}
 \quad, \label{second_int}
 \ee
and the second boundary condition  $y_s=y(z=D/2)$ can be expressed
as
 \be
 \frac{D}{2\lD}=\int^{y_s}_{y_m}\frac{{\rm d}w}{\sqrt{2\cosh{w}
 -2\cosh{y_{m}}}}
 \quad.\label{bc_2p13}
 \ee
The procedure to solve these  equations is similar to the one used
in the previous section. For given $D$, $\phi_0$,  $\alpha$, $\chi$
and $a$,  the profiles $\phi^\pm(z)$ and the surface value $\eta_s$
are calculated numerically and inserted into the free energy
expression, eq~\ref{e19}, $F_{\rm tot}[\phi^\pm(z);\eta_s,D]$.
Taking the derivative of $F_{\rm tot}$ with respect to $D$, or
equivalently using the contact theorem as for the counterion-only
system, the osmotic pressure, $\Pi(D)$ is obtained,
\be
 \Pi(D) = k_{B}T~(c_{m}^{+} + c_{m}^{-} - 2c_{0})=
\frac{k_{B}T}{a^3}(\phi_{m}^{+} + \phi_{m}^{-} - 2\phi_{0}).
\ee

\section{Results}

To better appreciate the role of ionic species in determining the
phase transition we first present results for counterions-only and
in the absence of hydration interactions. These results will already
clearly show the most important feature of the model, namely the
possible lamellar-lamellar transition. We then predict the effect of
added salt. Finally, we include the hydration contribution and
compare model results with experiments.

\subsection{Ion Dependent Lamellar-Lamellar Transition}

Our model contains three parameters: $\alpha$, $\chi$ and $a$. For
$a$ we use the linear size of the surfactant headgroup, typically in
the range of 7\AA\ to 9\AA\ with cross-sectional area $a^2$. For the
phenomenological constants $\alpha$ and, in particular, for $\chi$
there is no direct and accurate experimental measurement. However,
estimates consistent with experimental data (as discussed in
Secs.~IIIC and ~IV), will be used hereafter.  In our model, positive $\chi$
presents the possibility of lateral phase separation. In addition,
for $\alpha<0$, the counterions tend to stay associated with the
charged headgroup and reduce the surface charge density.

In Figure~3 we compare the osmotic pressure isotherm $\Pi(D)$ in the
case of no added salt, for three $\alpha$ values (and for a constant
$\chi=12$ and $a=8$\AA) with the standard PB isotherm (short
dashes), for the fully dissociated limiting case, $\eta_s=1$.
Formally, full dissociation can be achieved  by setting
$\alpha\to\infty$ in our equations.  For the two values of
$\alpha=-5.85,-5.95$, the isotherms in Figure~3 show a first-order
phase-transition in the range of $10 {\rm\AA} \lesssim D \lesssim
50$\,\AA. The  phase transition is from a dilute and highly charged
L$_\alpha$ lamellar phase (large $D$ and $\eta_s\lesssim 1$), to
another L$_{\alpha'}$ phase that is more condensed and less charged
(small $D$ and $\eta_s \ll 1$). As $\alpha$ increases, this phase
transition shifts to higher values of $\Pi$ and lower values of $D$.

\begin{figure}
\includegraphics[keepaspectratio=true,width=90mm,clip=true]{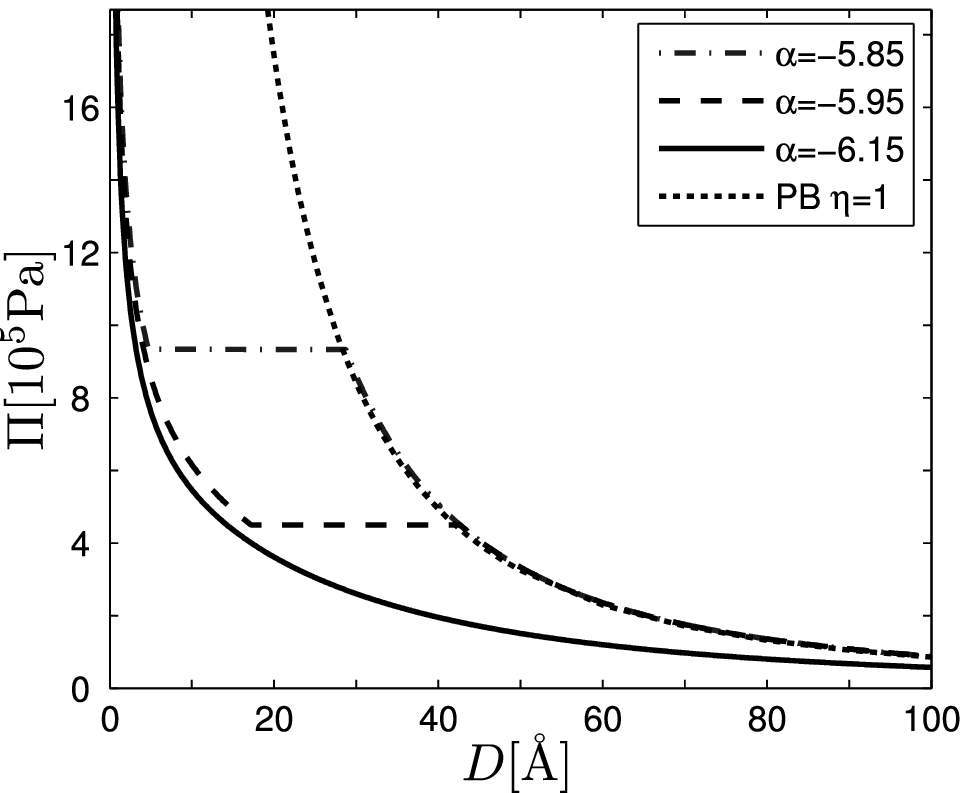}
  \caption{\footnotesize\textsf{The osmotic pressure isotherm $\Pi(D)$ for three
  binding strengths:  $\alpha=-5.85$ (dotted-dashes), $-5.95$ (dashed line), and  $-6.15$ (solid
  line). The other parameters are for non-ideal mixing
  $\chi=12$ and lateral separation $a=8$\,\AA. The $\alpha=-5.85$ and $-5.95$
  lines show a phase transition, while $\alpha=-6.15$ does not show one. For
  comparison,
  the usual PB isotherm with $\eta_s=1$ (short dashes) is also shown.}}
\end{figure}

For large $D$, $\eta_s$ is very close to one, and the osmotic
pressure isotherm closely follows the PB result; for large $D$ the
solution of our model remains essentially on the PB branch of the
free energy, characterized by  $\eta_s \simeq 1$. For small values
of inter-lamellar spacing, $D$, the values of osmotic pressure for
all three values of $\alpha$ are again very similar, with small and
slowly varying $\eta_s$, as in Figure~2b. Here, the system essentially
remains on the associated branch of the free energy characterized by
$\eta_s \ll 1$.

The lowest $\alpha$ isotherm (solid line, $\alpha=-6.15$)
shows no transition. The counterions are almost fully associated in
this case for the entire range of $D$, leading to a lower value of
pressure for all $D$'s. Note that all three osmotic pressure
isotherms as well as the PB one have the same limiting behavior for
$D\to\infty$. This is quite accurately described by the Langmuir
form of the osmotic pressure, as applied to the counterion-only case
 \cite{andelman95}:
\be
 \Pi(D)= \frac{k_B T  \pi}{2\lB}\frac{1}{D^2}~,
\ee
which does not depend on the value of the surface change (or,
equivalently on $\eta_s$).

In Figures~4 and 5 we show the effects of the variation of linear
size $a$ and lateral interaction $\chi$, respectively. We have
chosen the parameter range to show isotherms without a phase
transition (small $a$ or $\chi$) as well as those showing the
transition  (higher values of $a$ or $\chi$) in each of the figures.
The main features are the same as in Figure~3. The large $D$ region
represents highly dissociated lamellae (strongly charged), while the
phase transition (when it exists) can be seen for small $D$ at
higher lamellar density. Increasing $a$ moves the transition point
towards higher values of osmotic pressure or equivalently lower
values of $D$ until it eventually disappears. Increasing $\chi$ has
the same effect. Note also  that the pressure at low $D$ is
diminished on increase of $a$. This effect can be understood  by
recalling the relation $\sigma=e\eta_s/a^2$, so that for the same
$\eta_s$, larger $a$ corresponds to smaller surface charge density
$\sigma$.

\begin{figure}
\includegraphics[keepaspectratio=true,width=90mm,clip=true]{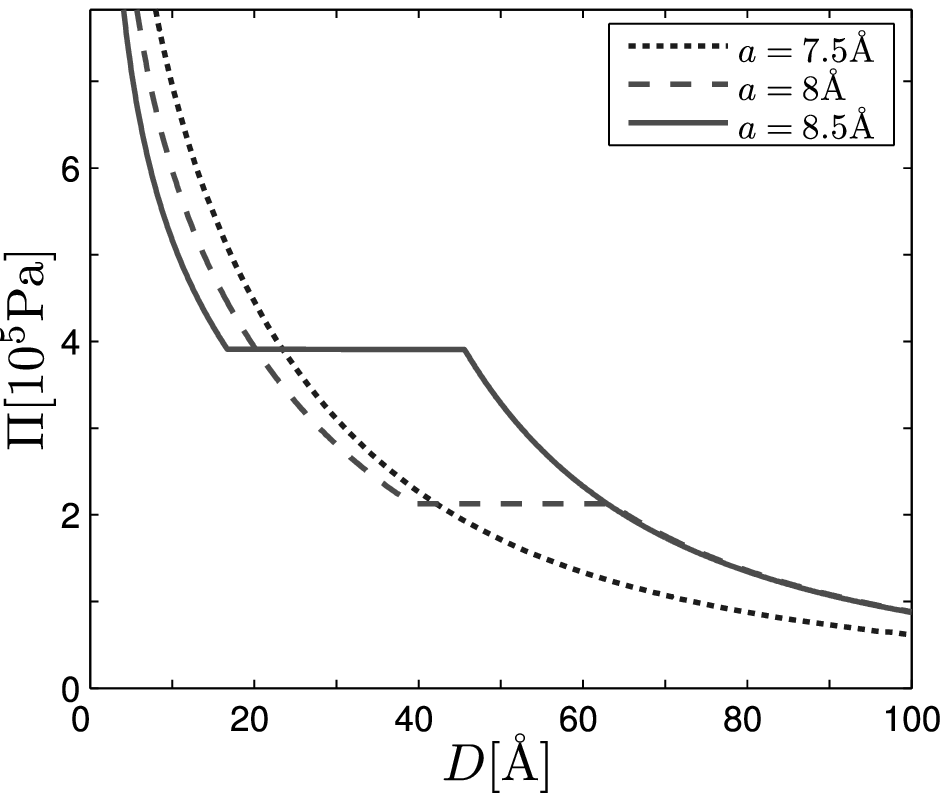}
 \caption{\footnotesize\textsf{The osmotic pressure isotherm $\Pi(D)$ for three
values of the surfactant headgroup separation:
  $a=7.5$\,\AA\ (short-dashes), $8$\,\AA\ (dashed line), and $8.5$\,\AA\ (solid line).
  The other parameters are
  $\alpha=-6$ and $\chi=12$. A
  phase transition is seen for $a=8$\,\AA\ and $8.5$\,\AA, but not for
  $a=7.5$\,\AA.}}
\end{figure}
\begin{figure}
\includegraphics[keepaspectratio=true,width=90mm,clip=true]{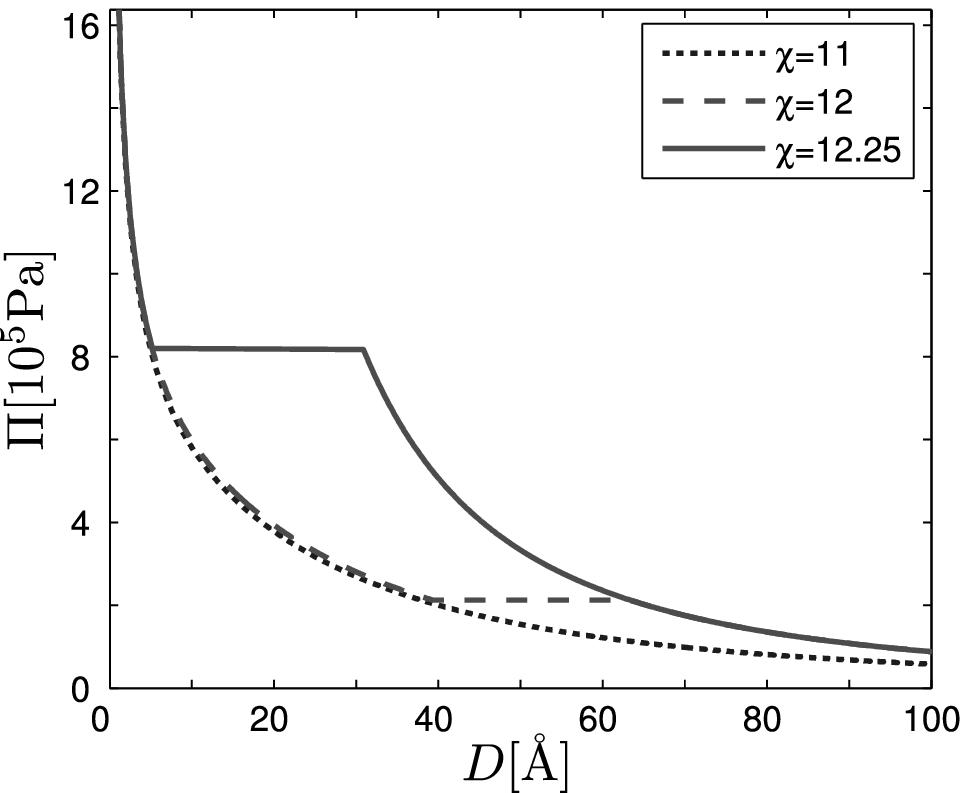}
 \caption{\footnotesize\textsf{The osmotic pressure isotherm $\Pi(D)$ for three
$\chi$
 values:
  $\chi=11$ (short dashes), $12$ (dashed line), and $12.25$ (solid line).
  The other parameters are $\alpha=-6$ and $a=8$\,\AA.
  The phase transition is seen for the two larger values of
 $\chi$. }}
\end{figure}

Our results are summarized in Figure~6 where the ($\alpha$, $\chi$)
parameter space is separated by a solid descending line
$\chi_c(\alpha)$ into two regions (for fixed $a$). The upper region
(large $\chi$, large $\alpha$) corresponds to isotherms with a phase
transition (and is designated as ``two phases" on the figure). Below
that region (small $\chi$, small $\alpha$) the isotherms show no
phase transition (designated as ``one phase" on the figure). The
degree of counterion dissociation varies in this region from very
small values to values $\eta_s\lesssim 0.8$ for finite values of
$\alpha$ and $\chi$. The PB result of $\eta_s=1$ is reached only
asymptotically as $\alpha\to \infty$. The line represents the
continuous line of critical point in the $(\chi,\alpha)$ plane. The
region between the solid and dot-dashed lines corresponds to jumps
in $D$ at the transition of more than $\approx 3$\,\AA. Above
the dot-dashed line, the behavior at $D\gtrsim 3$\,\AA\ is described
by the usual PB solution because the transition occurs at
unphysically small $D$ values.

\begin{figure}
\includegraphics[keepaspectratio=true,width=90mm,clip=true]{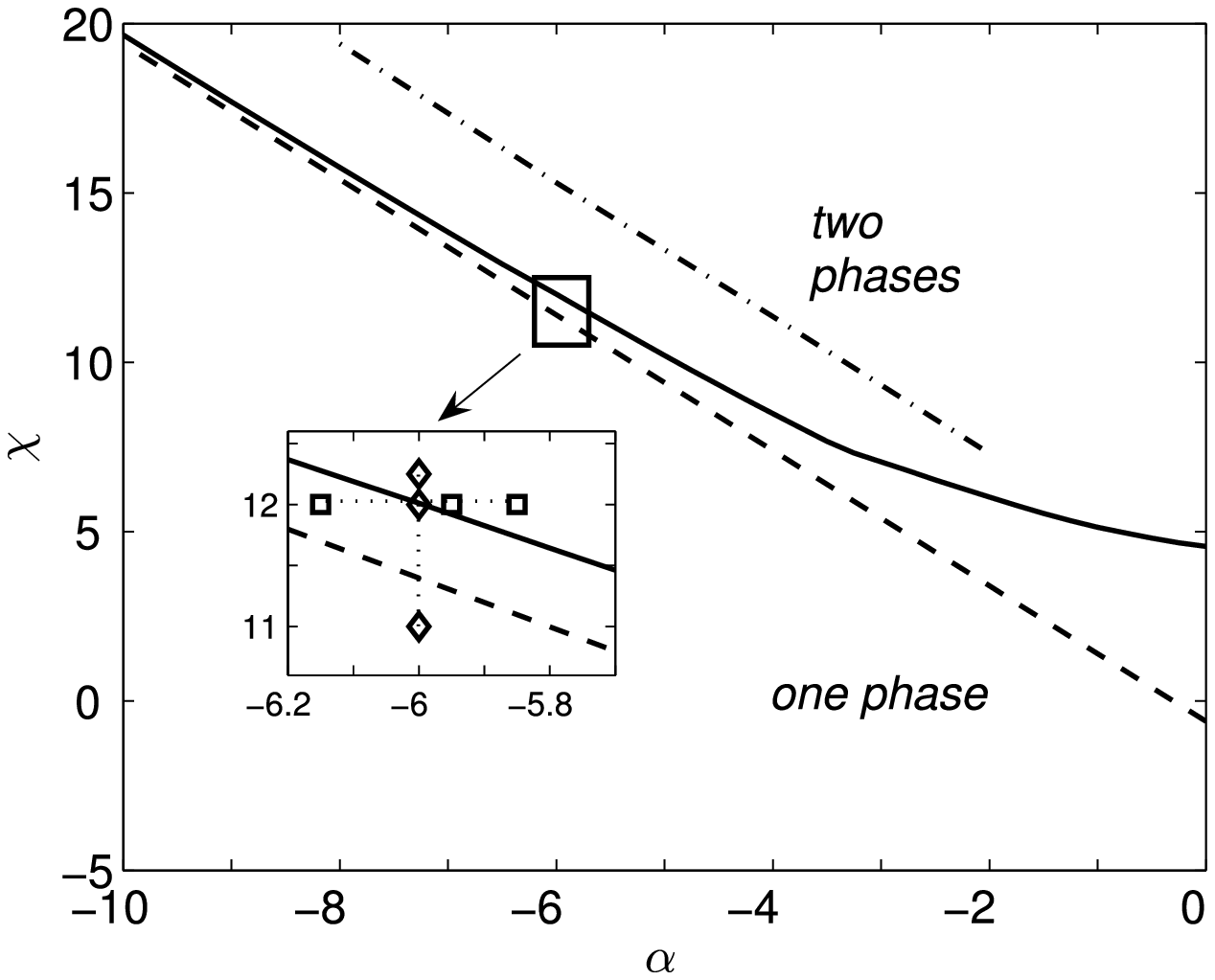}
  \caption{\footnotesize\textsf{Phase diagram   in
  the ($\alpha$, $\chi$) plane for $a=8$\,\AA. In the region below
  the solid line $\chi_c(\alpha)$, the system
  does not exhibit a
  phase transition (``one phase"). The dashed line is
  the analytical approximation of $\chi^*\simeq -2\alpha+2\ln(2\pi\lB/a)-4$.
  The inset is a blow-up of the region around $\alpha\approx -6$ and
  $\chi \approx 12$, showing the parameters used in figures 3 and 5.
  Square symbols correspond to isotherms in Figure~3, while diamonds correspond
to those in Figure~5. The region lying between the full and dot-dashed
lines in
  the two-phase region corresponds to transitions in D
  that are larger than $\approx 3$\,\AA. }}
\end{figure}

The inset to Figure~6 corresponds to variations of $\chi$ and
$\alpha$ shown in Figure~4 and 5, respectively. As
$\chi$ or $\alpha$ are lowered, the single-phase region is eventually reached.
Figure~6 is plotted for $a=8$\,\AA. As $a$ decreases, the two-phase
region shrinks and the one-phase region expands.

The dashed line in the figure is an analytic calculation which gives
the following approximate form of
$\chi^*(\alpha)$
\be \chi^*\simeq -2\alpha + 2\ln\left(\frac{2\pi\lB}{a}\right) -4
~.
 \ee
To derive this result we assume that the phase transition occurs at
large $D$. Using the asymptotic large $D$ behavior, we compare the
free energy of $\eta_s\simeq 1$ with $\eta_s\ll 1$ and determine the
transition point as a function of $\chi$ for given $\alpha$ and $a$.
As can be seen by comparing the analytic (dashed) line with the full
numerical solution (solid line), the approximation is good for small
$\alpha<0$. For $\alpha\gtrsim -3$ the assumption of a transition at
large $D$ breaks down and the approximated $\chi^*$ starts to
deviate considerably from the numerically calculated
$\chi_c(\alpha)$.

\subsection{Added Salt: Vanishing of the Transition}

The effect of added salt was treated in Sec.~IIc. The salt is
characterized by the Debye-H\"uckel screening length $\lD$, and
screens electrostatic interactions. As the amount of added salt
increases, $\lD$ decreases, and the phase transition observed in the
absence of salt becomes gradually less pronounced  until it is
finally wiped out completely~\cite{warr96}. 
This is clearly seen in Figure~7. In 7a,
three  osmotic pressure isotherms are shown. A plateau (first-order
phase transition) is seen for the two lower amounts of salt,
$c_b=10$ and 30\,mM, while the transition disappears for higher
amounts of salt, $c_b=50$\,mM.

\begin{figure}
 \includegraphics[keepaspectratio=true,width=80mm,clip=true]{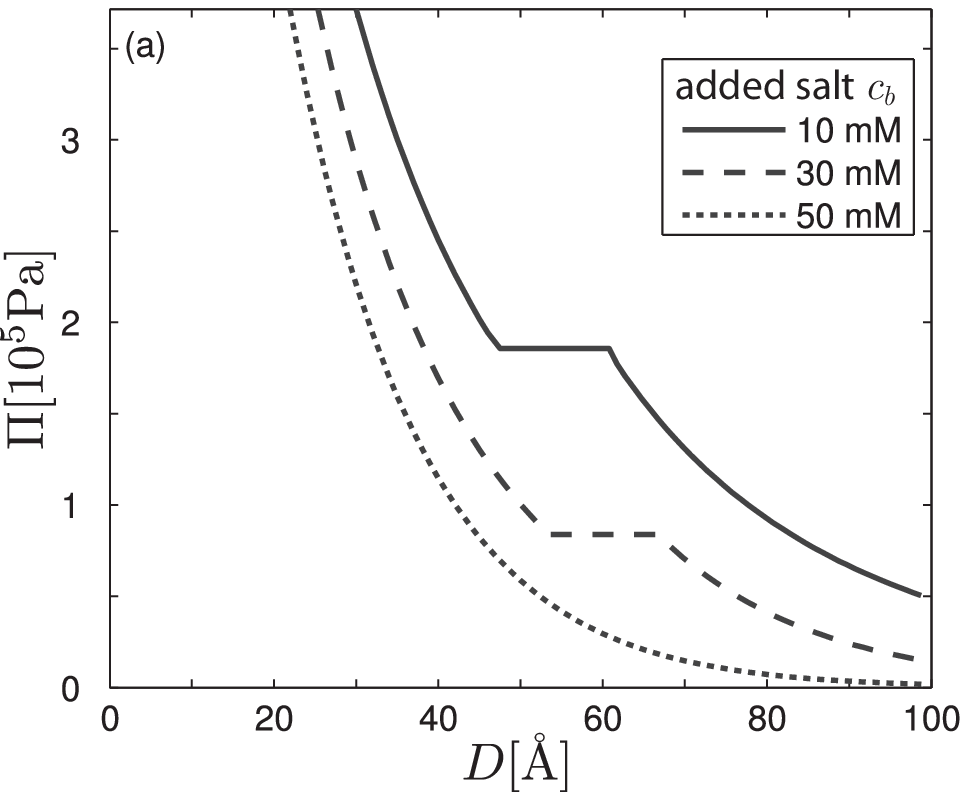}
   \includegraphics[keepaspectratio=true,width=80mm,clip=true]{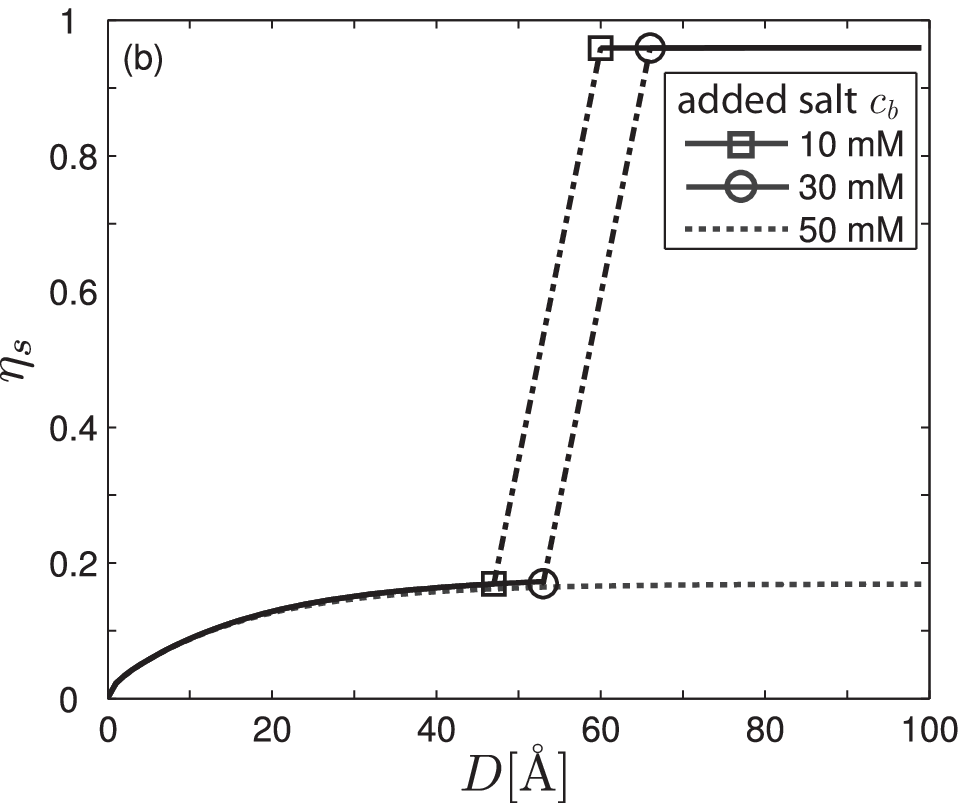}
  \caption{\footnotesize\textsf{Effect of added salt. In (a)
  $\Pi(D)$, in (b) $\eta_s(D)$
  are plotted as a function of added salt concentration
$c_{b}$.   In (a) and (b): $c_b=10$\,mM (solid line),
  $30$\,mM (dashed line), $50$\,mM (short dashes). In (c)
  salt concentration varies continuously over the range
  of existence of the transition, $0 \leq c_{b} \leq 40$\,mM. Other parameters
  are  $\alpha=-5$, $\chi=10.19$, and $a=8$\,\AA. As $c_b$ increases,
  screening becomes more important.  The entire osmotic pressure isotherm
$\Pi(D)$  decreases in magnitude and the phase transition region
  diminishes and shifts towards higher $D$ values.
  Note that for highest salt concentration,
  $c_b=50$\,mM,
  the  phase transition has vanished. In (b) the coexisting values
  of the two phases are denoted by a square ($c_b=10$\,mM) and by a circle
  ($c_b=30$\,mM). The dotted-dashed lines are the corresponding
  tie-lines.
   }}
\end{figure}

One can also see how the phase transition is first shifted towards
the high $D$ low $\Pi$ values, and then (for $c_b \approx 40$\,mM)
completely disappears. The overall decrease in $\Pi(D)$  as the
amount of salt increases, is due to the increased screening, and is
present also in the simple PB theory. In 7b the jump in $\eta_s$ is
shown for $c_b=10$ and 30\,mM, while it vanishes for higher amounts
of salt, $c_b=50$\,mM, in accord with the isotherm behavior.

It is instructive to follow the change of the transition pressure
$\Pi_{\rm tr}$ as salt is added to the system, which is related to the
difference in volume $\Delta V$ and number of ions $\Delta N$
in the two phases in a Clausius\--Clapeyron-like equation:
\be \frac{d \Pi_{\rm tr}}{d c_b}=\frac{k_BT}{c_b} \frac{\Delta
N}{\Delta V}. \label{neweq-1} \ee
 Remarkably, we find an almost linear dependence of
$\Pi_{\rm tr}$ in the whole range of $c_b$, starting with the
transition pressure at no added salt and leading eventually to the
loss of transition for sufficiently high salt concentrations, $c_b
\approx 40$\,mM, see figure 8. In eq~\ref{neweq-1}, $ d \Pi_{\rm
tr}/ d c_b $ corresponds to an added work of $2~k_{B}T $ due to the
exclusion of ions acting on the volume change at the transition.

\begin{figure}
   \includegraphics[keepaspectratio=true,width=80mm,clip=true]{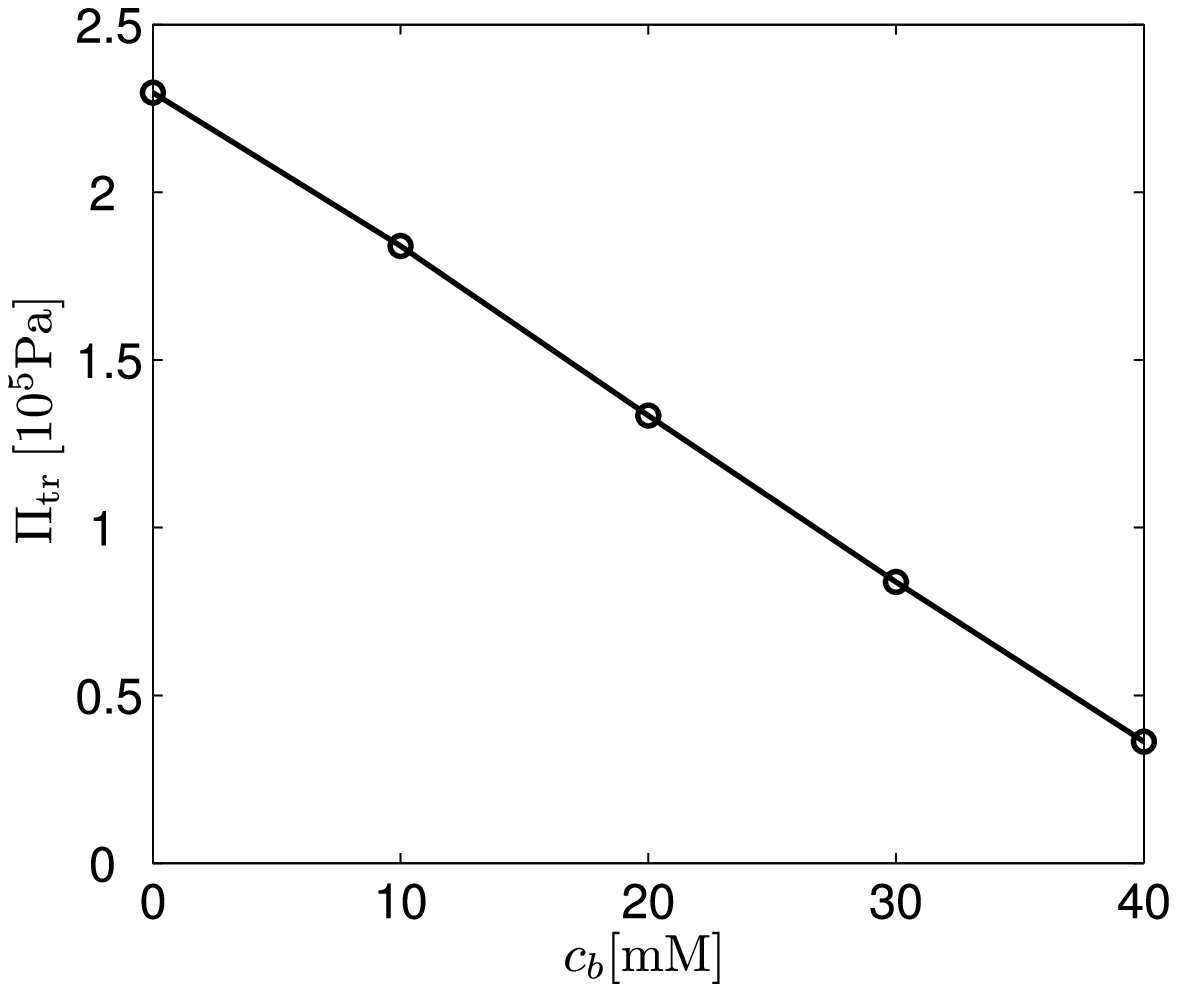}
  \caption{\footnotesize\textsf{
Effect of added salt on the transition pressure.
 $\Pi_{\rm tr}(c_{b})$ is plotted as a function of added salt
concentration $c_{b}$.
The dependence of the transition pressure on the salt
  concentration in the region where the transition
exists is linear, with a slope of $7.9 \times
10^{-21}$~J.
   }}
\end{figure}

It is also interesting to follow the change in $\Pi_{\rm tr}$ with
temperature.  Experimentally, both an increase of
$\Pi_{\rm tr}$ with $T$ (at lower $T$) and a decrease (at higher
$T$) have been observed \cite{zemb98}. In contrast, if we assume that
$\chi$ and $\alpha$ are independent of $T$ in the model, we find that $\Pi_{\rm
tr}$ monotonically, almost linearly, {\em increases} with $T$. This suggests
that in a more refined
model, the parameters $\chi$ and $\alpha$ should be taken as
functions of the temperature rather than simple constants. For
example, if we assume that
$\alpha(T)$ and $\chi(T)$ vary as $1/T$, while the ratio
$\alpha/\chi$ is kept constant, we find that $\Pi_{\rm tr}$
{\em decreases} monotonically with $T$.


\subsection{Relating to DDABr/DDACl Osmotic Pressure Experiments}


The experimentally observed difference between DDABr and DDACl for
the different halides can easily be rationalized within our model by
different values of  $\alpha$ and/or $\chi$, for the different ions.
This is indeed reasonable since experiments show that larger halide
ions have an added affinity even for neutral
lipids~\cite{tatulian83,rydall92,petrache05,zemb04}. The tendency of
ions to preferentially partition into the hydrocarbon-water
interface is most often reported in terms of an effective binding
interaction that acts in addition to the repulsive electrostatic
force, expected for ions interacting with low dielectric material.
These differences in binding affinity would  translate into a
different value of $\alpha$ within our model. In these terms,
experiments show that $|\alpha|$ is larger for bromide by one to
four $k_BT$ more than for chloride, and iodide is at least an order
of magnitude larger than those
\cite{tatulian83,rydall92,petrache05,zemb04}.

With our model assumptions, we can now try to fit the experimental
data in Ref.~\cite{zemb98} using the same (small) amount of added
salt as in the experiment, i.e.  $c_b=0.5$\,mM. The fit to the DDABr
and DDACl lamellar systems are shown in Figure~9. We will first
address the fit to the simpler case of DDACl that does not show in
experiments a liquid\---liquid coexistence, and then discuss DDABr,
where the liquid-liquid coexistence is clearly discerned.
%
The DDACl data was fitted in Figure~9a using $\alpha=-3.4$,
$\chi=14.75$ and $a=8$\,\AA. The DDACl data points, represented by
squares, are reproduced from Ref.~\cite{zemb98}. The value chosen for
$\chi$ is  higher, yet close to $\chi_c(\alpha)$ (See Figure~6).

\begin{figure}
\includegraphics[keepaspectratio=true,width=90mm,clip=true]{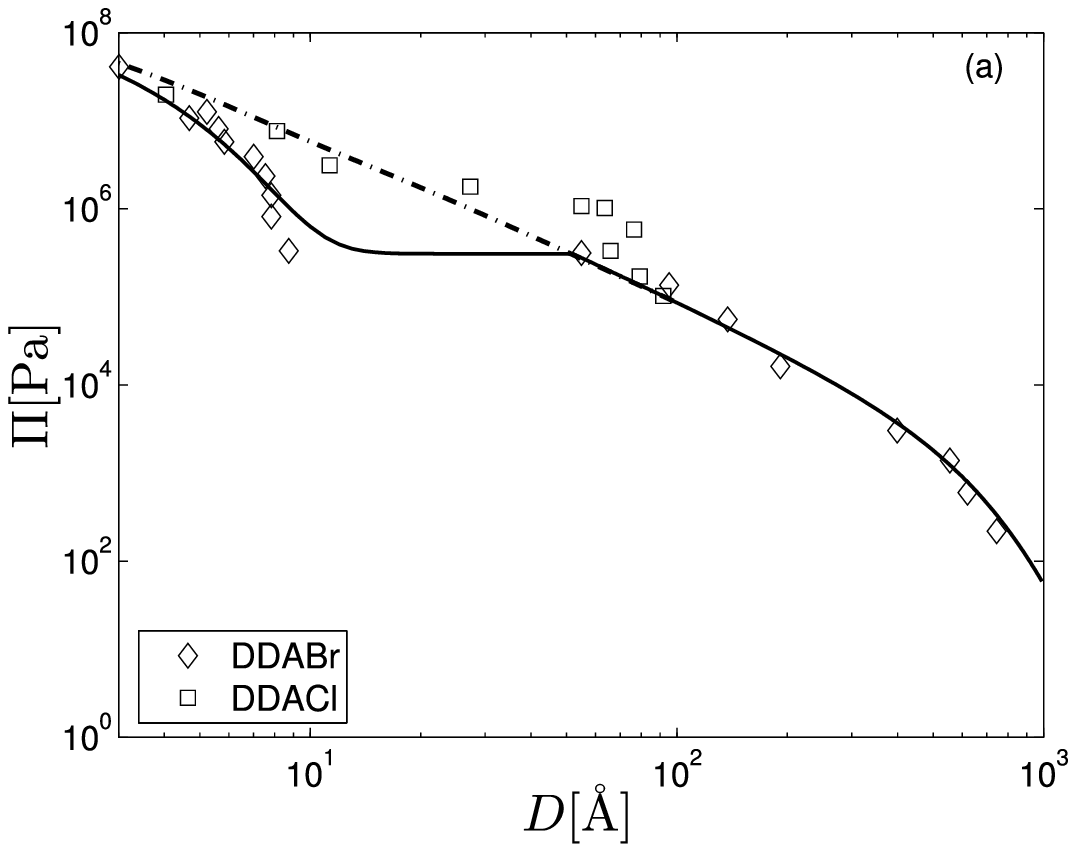}
   \includegraphics[keepaspectratio=true,width=90mm,clip=true]{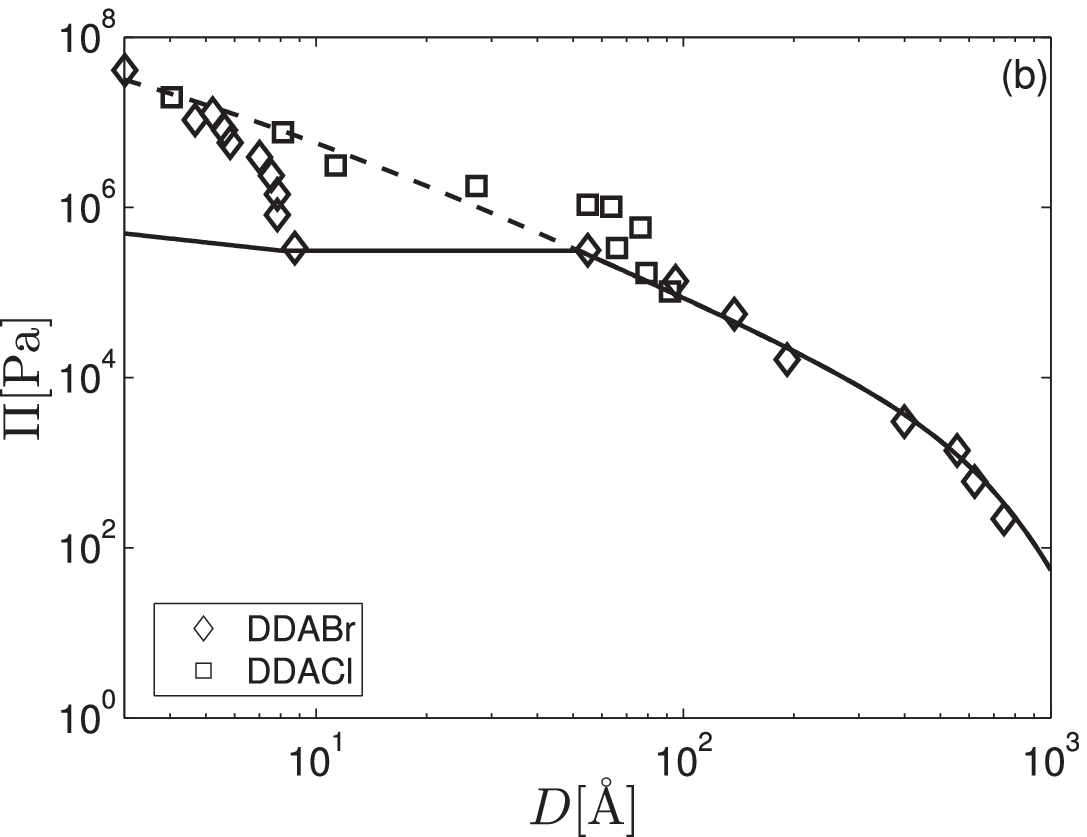}
  \caption{\footnotesize\textsf{Fit to the experimental osmotic pressure isotherm
$\Pi(D)$
  of Ref. \cite{zemb98} on a log-log scale.
  The diamonds and squares are the data points for DDABr and DDACl, respectively, reproduced
  from Ref.~\cite{zemb98}. In (a) the solid line is the best fit of the model to the phase
  transition seen for DDABr
  with $\alpha=-7.4$, $\chi=14.75$ and $a=8$\,\AA .  The fit also
  includes a hydration contribution (parameters for the form:
  $\Pi_{\rm hyd}=\Pi_0\exp(-D/\lambda_{\rm hyd})$, with typical values:
  $\Pi_0=2.37\cdot 10^8$\,Pa and $\lambda_{\rm hyd}=1.51$\,\AA).
  This contribution is particularly important at the low $D$ region of the
  DDABr isotherm. A small amount of salt is added in the fit
as in the experiment
  ($c_b=0.5$\,mM). The dot-dashed line is the fit to the DDACl (no
  transition).
  All parameters here are the same,
  except $\alpha=-3.4$ (one-phase region of Figure~6).
  In (b) the same model fits, showing only the electrostatic
  contributions to the force, $F_v+F_s$, that are responsible for the
  apparent phase transition.
  The fit to both data sets is good. However, the steep increase in pressure
  of the DDABr seen in the low $D$ range ($D<10$\,\AA) cannot be
  reproduced with these forces alone.
  }}
\end{figure}

In comparing to experiments, it is important to realize that for
small spacings, interactions of non-electrostatic origin tend to
dominate the osmotic
equilibrium~\cite{zemb98,petrache98,evans86,vap-hyd,petrache04}.
Even for highly charged systems, hydration interactions acting at
very short range, practically independent of the charging
equilibrium at the surface, invariably dominate at separations of
$D\lesssim10$\AA ~are certainly among the most important to
consider. Therefore, we add the hydration interaction $F_{\rm hyd}$
as an idealized separable term to the overall interaction energy:
$F_{\rm tot}+F_{\rm hyd}=F_v+F_s+F_{\rm hyd}$. In conformity with
hydration interactions between lipids \cite{parsegian89} and between
other macromolecules \cite{vap-hyd}, we use the phenomenological
form of an exponential interaction with a salt-independent decay
length, $\lambda_{\rm hyd}$ \cite{leikin}
\be
 F_{\rm hyd}=\Pi_0\lambda_{\rm hyd}\exp(-D/\lambda_{\rm hyd}).
\ee
To account for the hydration interaction, we fit with
$\Pi_0=2.37\times 10^{8}$\,Pa and $\lambda_{\rm hyd}=1.51$\,\AA.
These values are similar to those found from fits to experimental
inter-lamellar spacings of lipids\cite{vap-hyd,nagle00,zemb98}.

Note that our extended PB model predicts a transition at these
values of $\alpha$ and $\chi$. However, this phase transition occurs
at very small inter-lamellar separations of $D\lesssim 3$\AA,
indicating that only at such low $D$ values are the ions associated
(small $\eta_s$). For higher $D$ values, the ions dissociate and the
model follows the regular PB osmotic pressure. Values of $D\lesssim
3$\AA\ are well below the validity limit of the model; for such low
$D$'s the computed osmotic pressure isotherm can not be considered
realistic. For higher $D$, the surface charge density, $\eta_s$, is
almost one and the system exhibits the usual Poisson-Boltzmann
isotherm, with screened electrostatic interactions at large $D$. We
remark that the fit in the high $D$ region is very good, and
surprisingly, also persists to small $D$ (although there are very
few data points below $D=30-40$\,\AA).

The DDABr was fitted in Figure~9a using a different $\alpha=-7.4$
but the same $\chi=14.75$ and $a=8$\AA, as well as the same hydration
interaction.
The value of $\chi$ was
picked near $\chi_c$ so that the transition would occur at a value
of $D$ in the range of $10$ to $40$\,\AA. In this range our model is
still valid and nicely reproduces a coexistence between the two
liquid lamellar phases L$_\alpha$ and L$_{\alpha'}$.

To better appreciate the role of hydration, figure~9b shows the model
fits that exclude the hydration contribution to the free energy. At
smaller separations, $D<10$\AA, the fit deviates considerably from
the experimental osmotic pressure isotherm. While electrostatic forces
 can account
for the transition point itself, they can definitely not by themselves
reproduce the sharp rise in the osmotic pressure in the undissociated branch
following the phase transition, compare Figure~9a.

Juxtaposing Figures 9a and 9b, it is evident that adding the
hydration contribution affects mainly the undissociated branch at
small inter-surface separations, since at large spacings
electrostatic PB contributions completely overwhelm the much shorter
ranged hydration. The hydration interaction mainly affects the small
$D$ region of both DDACl and DDABr.
 In fact, while the magnitude of hydration
forces may not differ greatly from other (electrostatic) acting
forces, the rate of force decay (slope) at such low spacings
indicates that hydration is always acting. One can observe this in
the context of lipids \cite{petrache04} as well as DNA
\cite{rudi-DNA}. Fits to both DDACl and DDABr should, therefore,
contain hydration contributions to the free energy.

Importantly, as can be realized from Figure~6, there are
different ways to cross the
transition line, $\chi_c(\alpha)$, to witness the liquid\---liquid
coexistence. Another possible choice of parameters could thus assume
the variation in $\chi$ instead of $\alpha$. For example, the data
can be equally well fit if we fix $\alpha=-7.4$ and fit the DDABr
data with $\chi=14.75$, and the DDACl  with $\chi=18$.

Finally, we
address the iodide analog, DDAI, that in experiments
appears not to form a  swollen
liquid phase, remaining in the condensed liquid lamellar phase for
the whole range of osmotic pressure values~\cite{zembpc}. In the
language of our model DDAI has either large $\vert \alpha\vert$ or $\chi$
values; it never enters the PB branch but instead remains in the
undissociated branch for all spacings $D$. This corresponds to the
``one phase" region of figure~6. Indeed, $|\alpha|$ for iodide could
be estimated as yet larger than in the case of DDABr, probably by up
to an order of magnitude~\cite{tatulian83,zemb04}. The system
is confined to remain on the undissociated branch for this
counterion: the repulsion due to electrostatic interactions and
entropy of counter-ions is not strong enough to combat the (van der
Waals) attractions. These attractions that are strong enough to hold
the system in the secondary free energy minimum, effectively
prevent the swelling observed for the other two counterions. We
could thus establish a ranking in the model's parameter space, where
DDACl, DDABr and DDAI would make a monotonic Hofmeister-like series
either in the values of $\vert\alpha\vert$ or $\chi$. The 3rd
parameter, $a$, is not changed in the fit because it is taken as the
size of the same DDA$^+$ headgroup. 


\section{Discussion}


The model presented in this paper combines electrostatic and
non-electrostatic interactions between charged surfactant bilayers.
The non-electrostatic part of the interactions is accounted for
using two phenomenological parameters characterizing the strength of
the counterion-amphiphile interaction, $\alpha$, and lateral
amphiphile-amphiphile interactions, $\chi$, on the charged
dissociable surface. By choosing $\alpha$ to be negative,
corresponding to favorable adsorption energy, and $\chi$ to be
positive, corresponding to a net attractive interactions between
like-species, promoting lateral
segregation between the dissociated and non-dissociated surfactants
on the surface, we are able to qualitatively explain the
experimentally observed  ${\rm L}_\alpha \to {\rm L}_{\alpha'}$
lamellar-lamellar phase transition.

\subsection{Origin of Phase Transition}

An abundance of experimental
verification~\cite{tatulian83,rydall92,petrache05,zemb04} indicates
that the different halides interact differently with lipids, forcing
 us to recognize the existence of non-electrostatic
interactions at work. In our model, this preferential interaction is
represented by $\alpha$. By reducing the effective layer charge
density, a favorable preferential interaction of ions to the
interface acts to lower the pressure at any given spacing $D$.
However, $\alpha$ alone cannot account for the abrupt jump in $D$
associated with  a first-order phase transition. In fact, in the
absence of $\chi$, pressure isotherms for any value of $\alpha$ are
smooth (no phase transition).

Our model offers a natural extension of the PB theory with the
Langmuir-Frumkin-Davies adsorption theory as applied to simple
counterions. The large difference in behavior of the halide ions is
modeled by the parameters $\alpha$, $\chi$ and $a$. The
interaction parameters $\alpha$ and $\chi$ necessarily involve
contributions from changes in hydration, solvation and desolvation,
of lipid headgroups and their counterions. The model, therefore,
underscores the important role of water structuring around ions
devolved in the bulk versus at the interface.

The salient feature of our model is the first-order transition in
the osmotic pressure isotherm from an almost completely dissociated
state (highly charged and swollen, PB branch) at large
interlayer separations, to an almost neutral, weakly dissociated,
state approaching bilayer contact (condensed, undissociated branch).
The existence of this
transition depends on the values of both
phenomenological parameters, but it is present over an extended region of
phase space.  The PB branch of the osmotic pressure isotherm is not
much different from the standard PB theory with complete
dissociation, both with or without added salt. On the other hand,
the undissociated
branch is characterized by a large attenuation in the magnitude of
the osmotic pressure for a given inter-lamellar spacing, being about
two orders of magnitude smaller then in the PB case.

In our model the ${\rm L}_\alpha \rightarrow {\rm L}_{\alpha'}$
transition in the inter-lamellar spacing is coupled to a lateral
first-order phase transition of the $\eta_s$ order parameter. This
is a direct consequence of the coupling between inter-lamellar
electrostatic degrees of freedom (mean electrostatic potential, mean
ion density) and the surface non-electrostatic degrees of freedom as
quantified by the phenomenological parameters $\alpha$ and $\chi$.
The ensuing liquid\---liquid lamellar phase transition is thus not
only from one state of the lamellae where a larger fraction of the
amphiphiles are charged to another state where they are less
charged, but also from a state where the inter-lamellar forces are
by and large electrostatic in nature, to a state where they are
dominated by hydration. While in many experimental systems this
transition is smooth and gradual, it is quite pronounced and
discontinuous in the system studied here and in Ref.~\cite{zemb98}.

\subsection{Relating Model Parameters and Molecular Interactions}

 We propose that non-ideal mixing between counterion-associated
and dissociated surfactants can be responsible for an in-plane
transition, which, in turn, is coupled to the bulk transition. This
proposed non-ideality is represented in our model by $\chi$, as is
sometimes used to report on lipids showing phase transitions
following changes in pH \cite{blume00}. While at present direct
experimental verification and estimates for the proper $\chi$ values
are lacking, we propose that conformational changes of lipid
headgroups, ions and water structuring induced by the
adsorbing ion, together with an added van der Waals interaction
between neutralized surfactant complexes can lead to significant demixing.
Furthermore,
because larger ions are expected to perturb the lipid-water
interface to a greater extent, it is reasonable to expect that the
value of $\chi$ will follow a similar ranking to the binding of ions
to the interface, represented by $\alpha$.  We note, that the $\chi$
values needed to observe a phase transition, typically $\approx 10
k_B T$, are quite high. These high values are needed to overcome the
electrostatic repulsion between like-charged lipids in this unscreened,
highly charged system.
The source of this lipid demixing energy (our $\chi$
parameter) could be associated with mismatch of
 headgroup-headgroup interactions, such as
hydrogen bonding between neutral lipids, water-structuring forces,
 or non-electrostatic
ion-mediated interactions between lipids across two apposed bilayers
for small inter-lamellar separations.

The parameter $a^2$ models the area per headgroup on the membrane
plane. It is a function of several molecular interactions and, in
principle, can be determined variationally. In Ref.~\cite{zemb98}
the area/headgroup was found to vary in a non-trivial fashion, from
a larger value in the condensed lamellae to a smaller one in the
dilute lamellae. The forces determining the area per surfactant are
as yet unknown. Therefore, in the model we have not allowed for
changes in area per surfactant, but note that it is not inconsistent
to assume that the area per headgroup differs for the neutral vs.
charged surfactant. The expansion of lipid area upon condensation,
contrary to what is typically observed in phase transitions of
lipids, could be evidence for direct attraction between Br$^-$ ions
and the lipid hydrocarbon core, as suggested previously by Ninham
{\em et al} \cite{ninham04}. This point deserves further
investigation. Interestingly, however, as seen in Figure 4, the
model predicts that lipids with larger area per headgroup can show a
larger transition gap, as was found experimentally 
in the case of GM-3 ganglioside with
weakly adsorbing counterions \cite{cantupc}.

Another point that merits further investigation is the dependence of
the isotherms on temperature. Both $\alpha(T)$ and $\chi(T)$ are
complex functions of the temperature, with specific dependence that
cannot be obtained from our model. A change in temperature affects
the values of $\alpha(T)$ and $\chi(T)$ and can change the plateau
pressure values, as was measured and reported in Figure~9 of
Ref.~\cite{zemb98}. A better understanding of $\alpha(T)$ and
$\chi(T)$ may offer an explanation to the non-monotonic behavior of
the plateau pressure as function of temperature \cite{zemb98}.

Finally we point out that the lack of direct experimental evidence at
this time, particularly for $\chi$, limits out predictions to be mainly
qualitative. More specifically, we can offer only a
qualitative explanation for the strong difference in the behavior
for different counterions, namely: no transition, transition and no
stable swollen lamellar phase for the DDACl, DDABr and DDAI
amphiphilic systems, respectively. Therefore, the
fit to the data points shown on Figure~8 should be regarded as a
tentative explanation of the mechanism behind the observed phase
transition. For example, we cannot establish whether
 the main difference in ionic interactions with the surface are
properly characterized by the value of $\alpha$ as opposed to
$\chi$.

\section{Concluding Remarks}

Previously, phase transitions in DDABr lamellar systems have been
theoretically attributed to either an ion-dependent van der Waals
attraction between layers ~\cite{ricoul98,zemb92a}, or to a
strong-coupling effect between adsorbed ions, expected for surfaces
with high charge density ~\cite{netz05}. Here, we have shown
that it is possible to account for the phase transition assuming a
non-electrostatic interaction between ion-dissociated and ion-bound
surfactants. We suggest that this interaction is ion specific and,
hence, we offer an explanation for the different behavior seen for the
three halide counterions.

The large phenomenological parameters we have found in our own fits
of the data (see Fig. 9) as well as the large energetic terms
assumed in the other approaches ~\cite{ricoul98,zemb92a,netz05}, all
indicate that substantial attraction necessarily acts to overcome
the electrostatic repulsion between surfactants. 
The molecular origin of this large
energy has not yet been determined, and further experimental
verification of the different phenomenological parameters is
required. It will also be of interest to see if the lateral phase
transition underlying the swelling transition and its dynamical
evolution from one lamellar state to the other, can be observed
directly in experiments.

We hope that in follow-up studies, a more microscopic approach
will be able to shed light on the origin of the phase transition in
these charged lamellar systems and how they relate to specific
molecular details.

\section*{Acknowledgements}

We are indebted to Th. Zemb for numerous comments and suggestions.
We benefited from discussions with L. Belloni, H. Diamant, M.
Dubois, and H.I. Petrache. D.A. acknowledges the hospitality of the
LPSB/NICHD (NIH), where this work was completed and support from
the U.S.-Israel Binational Science Foundation (B.S.F.) under grant
No. 287/02, and the Israel Science Foundation under grant No. 160/05. This
research was supported in part by the Intramural Research Program of
the NIH, NICHD.

\newpage


\newpage



\begin{thebibliography}{99}

\bibitem{parsegian93}
V. A. Parsegian, Langmuir {\bf 9}, 3625 (1993).

\bibitem{helfrich78}
W. Helfrich, Naturforsch. {\bf 33a}, 305 (1978).

\bibitem{evans86}
E. A. Evans, and  V. A. Parsegian, Proc. Natl. Acad. Sci. (USA)
{\bf 83} 7132 (1986).

\bibitem{andelman95}
D. Andelman, in \textit{Handbook of Biological Physics: Structure
and Dynamics of Membranes"},  Vol. 1B, edited by  Lipowsky R.;
Sackmann E. (Elsevier Science B.V., Amsterdam, 1995), ch. 12.

\bibitem{petrache98}
H. I. Petrache, N.  Gouliaev, S. Tristram-Nagle, R. Zhang, R. M.
Suter, and J. F. Nagle, Phys. Rev. E {\bf 57}, 7014 (1998).

\bibitem{zemb92a}
Th. Zemb, L. Belloni, M. Dubois, and S. Marcelja, Prog. Coll.
Polym. Sci. {\bf 89}, 33 (1992).

\bibitem{LesHouches}
C. Holm, P. K\'{e}kicheff, and R. Podgornik, (Editors),
\textit{Electrostatic Effects in Soft Matter and Biophysics} (
Kluwer,Dordrecht, 2001).

\bibitem{safinya89}
C. R. Safinya, E. B. Sirota, D. Roux, and
G. S. Smith, Phys. Rev. Lett. {\bf 62}, 1134 (1989).

\bibitem{guttman93}
G. D. Guttman, and D. Andelman, J. Phys. II (France) {\bf 3}, 1411 (1993).

\bibitem{harries03}
D. Harries, S. May, and A. Ben-Shaul, J. Phys. Chem. B
  {\bf 107}, 3624 (2003).

\bibitem{may02}
S. May, D. Harries, and A. Ben-Shaul,
 Phys. Rev. Lett.  {\bf 89} 268102 (2002).

\bibitem{arnold95}
K. Arnold, in \textit{Handbook of Biological Physics: Structure
and Dynamics of Membranes},  Vol. 1B, edited by R. Lipowsky R.;
Sackmann, E. (Elsevier Science B.V., Amsterdam, 1995) ch. 19.

\bibitem{parsegian71}
B. W. Ninham, and V. A. Parsegian, J. Theor. Biol. {\bf 31}, 405 (1971).

\bibitem{zemb98}
M. Dubois, Th. Zemb, N. Fuller, R. P. Rand, and V. A. Parsegian,
J. Chem. Phys. {\bf 108}, 7855 (1998).

\bibitem{zembpc} Zemb, Th. personal communication.

\bibitem{hofmeister}
F. Hofmeister, Archiv. Exp. Path.  Pharm. {\bf 24}, 247 (1887).

\bibitem{ninham04}
P. Kunz, P. L.  Nostro, and B. Ninham, Curr. Opin.
  Colloid Interface Sci. {\bf 9}, 1  (2004).

\bibitem{washabaugh85}
K. Collins, and M. Washbaugh, Q. Rev. Biophys. {\bf 18}, 323 (1985).

\bibitem{garrett04}
B. C. Garrett, Science {\bf 303}, 1146 (2004).

\bibitem{ninham97}
B. W. Ninham, and V. Yaminsky, Langmuir {\bf 13}, 2097 (1997).

\bibitem{attard88}
P. Attard, D. J. Mitchell, and B. W. Ninham, J. Chem. Phys. {\bf 88},
4987 (1988).

\bibitem{ninham03}
A. Becheri, P. L. Nostro, B. W. Ninham, and P. Baglioni, J.
  Phys. Chem. B {\bf 107}, 3979 (2003).

\bibitem{tobias02}
P. Jungwirth, and D. J. Tobias, J. Phys. Chem. B {\bf 106}, 6361 (2002).

\bibitem{gurau04}
M. C. Gurau, S.-M. Lim, E. T. Castellana, F. Albertorio, S. Kataoka,
and P. S. Cremer, J. Am. Chem. Soc. {\bf 126}, 10522 (2004).

\bibitem{tatulian83}
 S. A. Tatulian, Biochim. Biophys. Acta  {\bf 736} 189 (1992).

\bibitem{rydall92}
J. R. Rydall, and P. M. Macdonald, Biochemistry {\bf 31}, 1092 (1992).

\bibitem{petrache05}
 H. I. Petrache, I.  Kimchi, D.  Harries, and V. A.  Parsegian,
 J. Am. Chem. Soc. {\bf 127}, 11546 (2005).

\bibitem{zemb04}
Th. Zemb, L. Belloni, M. Dubois, A. Aroti, and E.
Leontidis, Curr. Opin. Coll. Int. Sci. {\bf 9}, 74 (2004).

\bibitem{sachs03}
J. N. Sachs, and T. B. Woolf, J. Am. Chem. Soc. {\bf 125}, 8742 (2003).

\bibitem{blume00}
P. Garidel, C. Johann, and A. Blume, J. Lip. Res. {\bf 10}, 131 (2000).

\bibitem{noro99}
M. G. Noro, and W. M. Gelbart, J. Phys. Chem. {\bf 111}, 3733 (1999).

\bibitem{nardi98}
J. Nardi, R. Bruinsma, and E. Sackmann, Phys. Rev. E {\bf 58}, 6340
(1998).

\bibitem{feigenson93a}
J. Huang, and G. W. Feigenson, Biophys. J. {\bf
    65}, 1788 (1993).

\bibitem{feigenson93b}
J. Huang, J. E. Swanson, A. R. G. Dibble,
  A. K. Hinderliter, and G. W. Feigenson, Biophys. J. {\bf 64}, 413
  (1993).

\bibitem{pabst05}
B. Pozo Navas, K. Lohner, G. Deutsch, E. Sevcsik, K. A. Riske,
R. Dimova, P. Garidel, G. Pabst,
Biochim Biophys. Acta {\bf 1716}, 40 (2005).


\bibitem{rand03}
N. Fuller, C. R. Benatti, and R. P. Rand, Biophys. J. {\bf 85}, 1667 (2003).

\bibitem{hill60}
Hill, T. L. \textit{Introduction to Statistical Thermodynamics}
(Addison-Wesley, New-York, 1960).

\bibitem{adamson90}
A. W. Adamson, and A. P. Gast \textit{Physical Chemistry of Surfaces}
 (Wiley and Sons, New-York, 1997), Chap. XI, XVI.

\bibitem{davis58}
J. T. Davies, Proc. Royal Soc. A {\bf 245}, 417 (1958).

\bibitem{andelman96}
H. Diamant H., and D. Andelman, J. Phys. Chem. {\bf 100} 13732 (1996);
{\em ibid.} Europhys. Lett. {\bf 34} 575 (1996).

\bibitem {andelman01}
H. Diamant, G. Ariel, and D. Andelman, Coll. Surf. A {\bf 183-185}, 259 (2001).

\bibitem{podgornik89}
R. Podgornik, Chem. Phys. Lett. {\bf 163}, 531 (1989).

\bibitem{kornishev92}
A. A. Kornyshev, D. A. Kossakowski, and S. Leikin, J. Chem.
Phys. {\bf 97}, 6809 (1992).

\bibitem{podgornik95}
R. Podgornik, and V. A. Parsegian, J. Phys. Chem. {\bf 99},  9491
(1995).

\bibitem{warr96}
H. N. Patrick, and G. G. Warr, J. Phys. Chem. {\bf 100}, 16268 (1996).

\bibitem{vap66}
V. A. Parsegian, Trans. Faraday Soc. {\bf 62}, 848 (1966).

\bibitem{vap-hyd}
V. A. Parsegian, R. P. Rand, D. C. Rau, Method Enzymol. {\bf 259}, 43
(1995).

\bibitem{petrache04}
H. I. Petrache, S. Tristram-Nagle,
K. Gawrisch, D. Harries, V. A. Parsegian, and J. F. Nagle,
Biophys. J. {\bf 86}, 1574 (2004).

\bibitem{parsegian89}
R. P. Rand, and V. A. Parsegian, Biochim. Biophys. Acta {\bf
988}, 351 (1989).

\bibitem{leikin}
S. Leikin, V. A. Parsegian, D. C. Rau, and R. P. Rand, Annu. Rev.
Phys. Chem. {\bf 44}, 369 (1993).

\bibitem{nagle00}
J. F. Nagle, and S. Tristram-Nagle,
  Biochim. Biophys. Acta Rev. {\bf 1469}, 159 (2000).

 \bibitem{rudi-DNA}
P. L. Hansen, R. Podgornik, and V. A. Parsegian, Phys. Rev. E.
{\bf 64}, 021907 (2001).

\bibitem{cantupc}
 P. Brocca, E. del Favero, and L. Cantu,
personal communication.

\bibitem{ricoul98}
F. Ricoul, M. Dubois, L. Belloni, Th. Zemb, C.
   Andr\'e-Barr\'es, and I. Rico-Lattes, Langmuir
  {\bf 14}, 2645 (1998).

\bibitem{netz05}
H. Boroudjerdi, Y.-W. Kim, A. Naji, R. R. Netz,
  X. Schlagberger, and A. Serr, Phys. Rep. {\bf 416}, 129 (2005).

\end{thebibliography}
\end{document}